%% file: main.tex
\documentclass[a4paper,11pt]{article}
\pdfoutput=1 

\usepackage{jinstpub} 
\usepackage{graphicx} 
\graphicspath{{fig/}}

\usepackage[utf8]{inputenc}
\usepackage{siunitx}
\usepackage{todonotes}

\title{Beam Test Performance Studies of CMS Phase-2 Outer Tracker Module Prototypes}

\newcommand{\pt}{$p_\textrm{T}$}

\newcommand{\micron}{\micro\meter}
\newcommand{\GEANTfour}{\textsc{GEANT}$4$}
\DeclareSIUnit\barn{b}
\DeclareSIUnit[quantity-product = ]\percent{\char`\%}
\collaboration[a]{The Tracker Group of the CMS Collaboration}

\emailAdd{oliver.pooth@cern.ch}

\abstract{
 A new tracking detector will be installed as part of the Phase-2 upgrade of the CMS detector for the high-luminosity LHC era.
  This tracking detector includes the Inner Tracker, equipped with silicon pixel sensor modules, and the Outer Tracker, consisting of modules with two parallel stacked silicon sensors. The Outer Tracker front-end {ASICs} will be able to correlate hits from charged particles in these two sensors to perform on-module discrimination of transverse momenta (\pt). 
  The \pt{} information is generated at a frequency of $\qty{40}{MHz}$ and will be used in the Level-1 trigger decision of {CMS}.
  Prototypes of the so-called {2S} modules were tested at the Test Beam Facility at {DESY} Hamburg between 2019 and 2020.
  These modules use the final front-end {ASIC}, the {CMS} Binary Chip ({CBC}), and for the first time the Concentrator Integrated Circuit ({CIC}), optical readout and on-module power conversion. In total, seven modules were tested, one of which was assembled with sensors irradiated with protons.
  An important aspect was to show that it is possible to read out modules synchronously.  
  A cluster hit efficiency of about $\qty{99.75}{\percent}$ was achieved for all modules.
  The {CBC} \pt{} discrimination mechanism has been verified to work together with the {CIC} and optical readout.
  The measured module performance meets the requirements for operation in the upgraded CMS tracking detector.
}


\begin{document}

 \maketitle

\input{DN-20-014_introduction.tex}
\input{DN-20-014_2Smodules.tex}
\input{DN-20-014_Setup-and-DAQ.tex}
\input{DN-20-014-Reconstruction_dataanalysis.tex}
\input{DN-20-014_results.tex}

\input{DN-20-014_conclusion.tex}

\acknowledgments
The measurements leading to these results have been performed at the Test Beam Facility at {DESY} Hamburg (Germany), a member of the Helmholtz Association {(HGF)}.
We also thank the team at the Karlsruhe irradiation facility for their support.

The tracker groups gratefully acknowledge financial support from the following funding agencies: BMWFW and FWF (Austria); FNRS and FWO (Belgium); CERN; MSE and CSF (Croatia); Academy of Finland, MEC, and HIP (Finland); CEA and CNRS/IN2P3 (France); BMBF, DFG, and HGF (Germany); GSRT (Greece); NKFIH K143477 and VLAB at HUN-REN Wigner RCP (Hungary); DAE and DST (India); INFN (Italy); PAEC (Pakistan); SEIDI, CPAN, PCTI and FEDER (Spain); Swiss Funding Agencies (Switzerland); MST (Taipei); STFC (United Kingdom); DOE and NSF (U.S.A.). This project has received funding from the European Union’s Horizon 2020 research and innovation programme under the Marie Sk\l odowska-Curie grant agreement No 884104 (PSI-FELLOW-III-3i). Individuals have received support from HFRI (Greece).


\newpage
\bibliographystyle{JHEP}
\bibliography{DN-20-014.bib}

\newpage
\include{TrackerAuthorList_2024_OT2STestbeam}

\end{document}

%% file: DN-20-014_introduction.tex
\section{Introduction} \label{sec:introduction}

The Large Hadron Collider (LHC)~\cite{Evans_2008} at CERN will be upgraded to the High-Luminosity {LHC} {(HL-LHC)}~\cite{Apollinari:2116337} by the year 2029 to deliver instantaneous luminosities of up to $\qty{7.5E34}{cm^{-2}s^{-1}}$.
The increase in luminosity places new demands on the experiments to maintain performance.
While the {Compact Muon Solenoid} ({CMS}) detector~\cite{Chatrchyan:2008aa, cmscollaboration2023development} was initially designed for instantaneous luminosities of $\qty{1E34}{cm^{-2}s^{-1}}$ and an average of $20$ to $30$~proton-proton collisions per bunch crossing (pileup), the pileup will increase to $140$ to $200$~collisions per bunch crossing at the {HL-LHC}.
Therefore, the {CMS} detector is being upgraded~\cite{CMSCollaboration:2015zni} as part of the Phase-2 upgrades.

\begin{figure}[!tb]
    \centering
    \includegraphics[width=\textwidth]{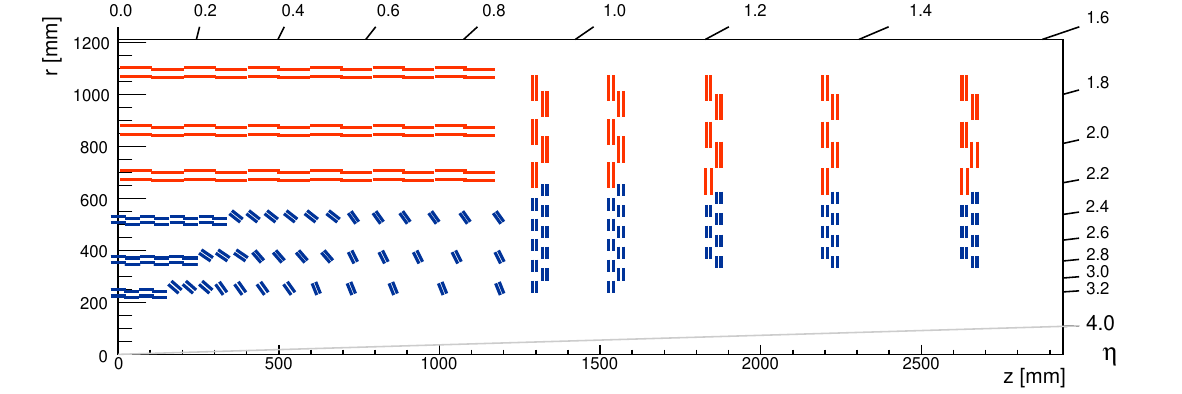}
    \caption{Sketch of one half of the CMS Phase-2 Outer Tracker layout in $r$-$z$ view~\cite{tkLayout}. The blue and red lines represent the two types of modules, PS modules equipped with one macro-pixel sensor and one strip sensor and 2S modules equipped with two strip sensors, respectively.\label{fig:tracker-layout}}
\end{figure}

The Phase-2 {CMS} detector will be equipped with a completely new silicon tracker~\cite{Klein:2017nke} consisting of a silicon pixel detector (the Inner Tracker) located close to the beam line and the Outer Tracker surrounding the Inner Tracker.
The layout of the Outer Tracker is shown in Figure~\ref{fig:tracker-layout}.
The tracker is designed to be radiation tolerant up to the integrated luminosity of $\qty{4000}{\femto\barn^{-1}}$ expected after ten years of HL-LHC operation, and to provide enhanced granularity for efficient particle tracking at the increased pileup.
In addition, the new silicon tracker will have a lower material budget than the current tracking detector.
For the first time, information from the silicon tracking system will contribute to the {CMS} Level-1~(L1) trigger decision~\cite{CERN-LHCC-2020-004}.
Two different types of modules are used for the Outer Tracker: PS modules in the region $\qty{217}{mm} < r < \qty{611}{mm}$, equipped with one macro-pixel sensor and one strip sensor, 
and 2S modules in the region $\qty{614}{mm} < r < \qty{1100}{mm}$, equipped with two strip sensors.

\begin{figure}[!tb]
  \centering
  \includegraphics[width=.65\textwidth]{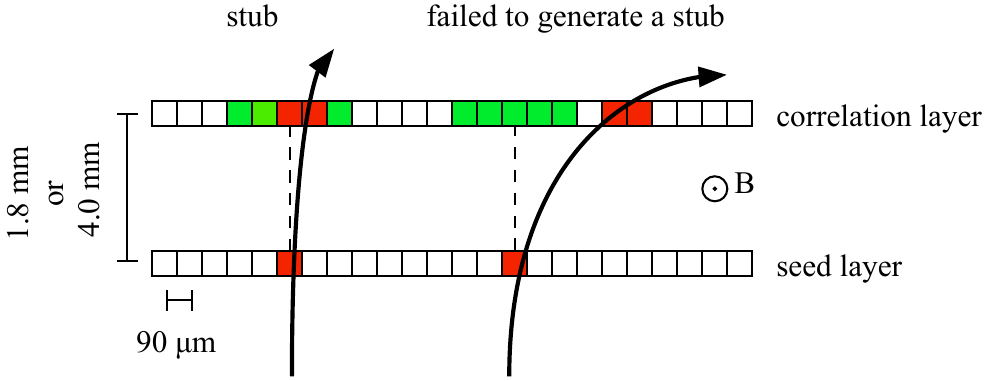}
  \caption{Sketch to visualize the stub formation with two parallel silicon strip sensors in a magnetic field.
  Silicon strips oriented into the viewing plane with a pitch of $\qty{90}{\micro\meter}$ are illustrated by boxes.
  The detected clusters are shown in red. The programmable search window is illustrated in green.
  The left arrow shows a high \pt{} and the right arrow a low \pt{} traversing charged particle.
  \label{fig:stub-mechanics}}
\end{figure}

To cope with bandwidth constraints, one of the key features of the Phase-2 Outer Tracker is the discrimination of the transverse momentum \pt{}. 
Both module types will provide information for the L1 trigger at the LHC bunch crossing frequency of $\qty{40}{MHz}$, based on the transverse momentum of passing charged particles, whose trajectories are bent in the $\qty{3.8}{T}$ magnetic field of the CMS detector.
Full event information containing all particle hits independent of the transverse momentum is sent out only after the reception of an L1 trigger accept signal, at a maximum average rate of $\qty{750}{kHz}$.
By using modules with two stacked silicon sensors in the magnetic field of the CMS detector, the \pt{} information can be inferred by combining the hit information from the two sensors in the front-end electronics.
Figure~\ref{fig:stub-mechanics} shows a sketch of the stub mechanism for 2S modules.
When a cluster is detected in the first sensor (seed layer), the closest cluster within a programmable search window on the second sensor (correlation layer) generates a cluster pair.
Stubs are formed from cluster pairs compatible with particles above the chosen \pt{} threshold, which is determined by the search window size.
By applying a \pt{} threshold of $\qty{2}{GeV}$ the amount of data can be reduced by about $\qty{90}{\percent}$.
The presence of particles with higher \pt{} is an indication of a hard scatter and thus that a potentially interesting physics process has taken place.

Seven prototype {2S~modules} were exposed to an electron beam at the DESY Test Beam Facility~\cite{Diener:2018qap} between December 2019 and November 2020.
One of the modules was built from sensors irradiated with $\qty{23}{MeV}$ protons~\cite{ETPIrrad} up to a $1$-MeV neutron equivalent fluence of $4.6 \times 10^{14}\,\textrm{n}_{\textrm{eq}}\textrm{cm}^{-2}$.
This fluence corresponds to $\qty{91}{\percent}$ of the maximum fluence expected for {2S~modules} after ten years of HL-LHC operation with an integrated luminosity of $\qty{4000}{fb^{-1}}$.
Data from the irradiated sensors shown in this paper were gathered after an equivalent annealing time of 200~days at room temperature, which corresponds roughly to the expected
annealing state at the end of the HL-LHC operations.
Further details about the irradiated module can be found in Ref.~\cite{PhdRolandKoppenhoefer}. 
This was the first time that full-size prototypes of {2S~modules} were optically read out in a particle beam.
Another important aspect of the test beam was to demonstrate the ability to read out multiple modules synchronously in a particle beam for the first time, validating recent software and firmware developments.
This paper summarizes the results of efficiency and resolution analyses performed with data of these test beam campaigns.
Results from beam tests with earlier prototypes are reported in Refs.~\cite{Adam:2018krk,Adam:2020dgr, LastBTPaper}.

%% file: DN-20-014_2Smodules.tex
\section{{2S} Modules for the {CMS} Phase-2 Outer Tracker} \label{sec:2Smodules}

\begin{figure}[!tb]
  \centering
  \includegraphics[width=.87\textwidth]{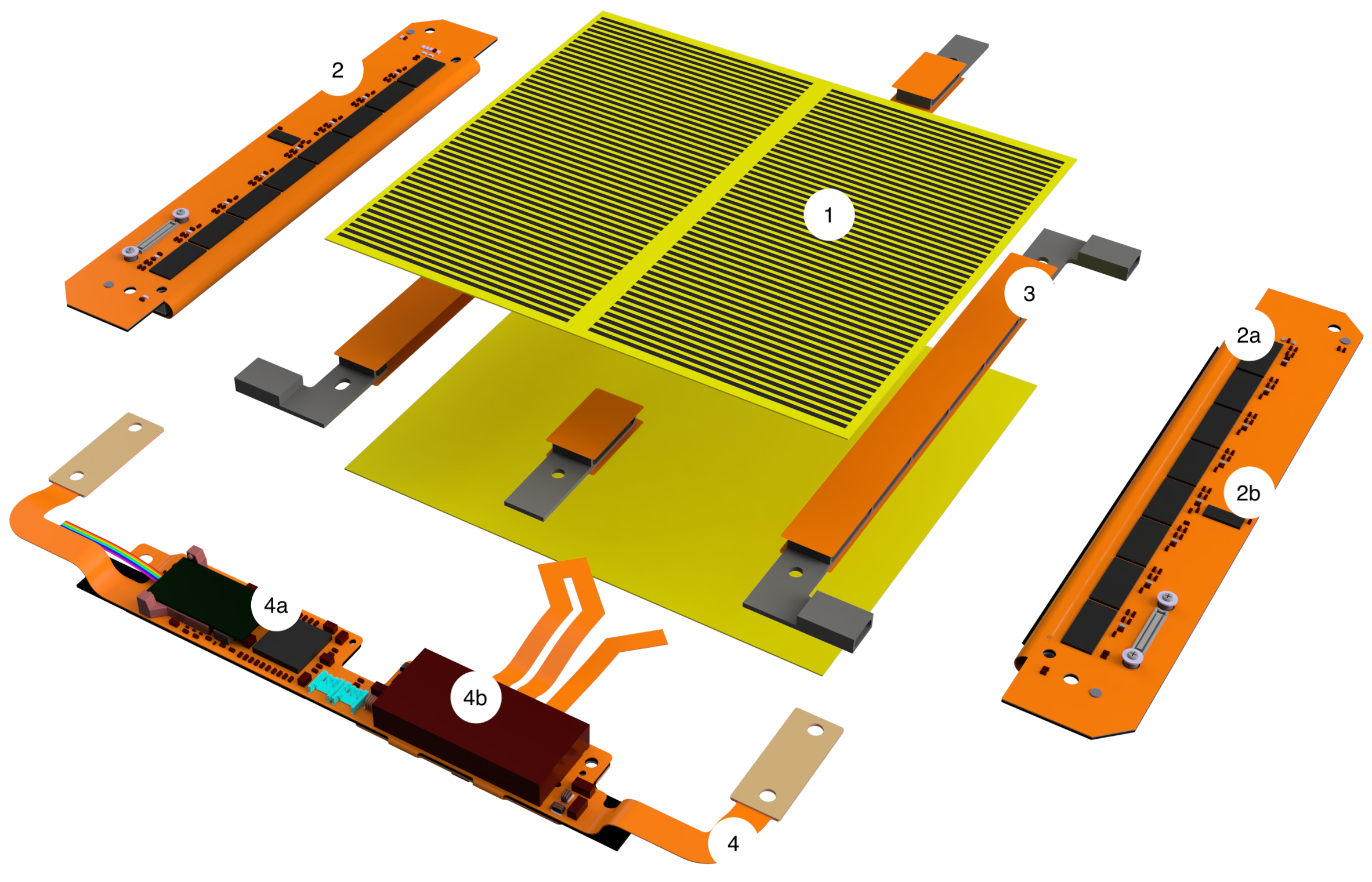}
  \caption[Sketch of a {2S} module in exploded view]{The {2S} module, shown as an exploded view, consists of two identical silicon strip sensors (top sensor and bottom sensor), with lines representing the strips for illustration~(1), two front-end hybrids (2), each with eight CMS Binary Chips~(2a) and a  Concentrator Integrated Circuit for data compression~(2b), and Kapton-isolated carbon fiber reinforced aluminum bridges~(3) separating the two sensors. On the front side a service hybrid~(4) with a Versatile Link Plus Transceiver, a low-power Gigabit Transceiver~(4a), and low-voltage DC-DC converters below a shield~(4b) completes the module.}\label{fig:2s-module-explosion-sketch}
\end{figure}

A {2S~module} consists of two parallel silicon strip sensors, readout electronics, service electronics, and mechanical parts.
An exploded view of the module is shown in Figure~\ref{fig:2s-module-explosion-sketch}.
The n-in-p type, $\qty{103}{\milli \meter}$ by $\qty{94}{mm}$ large, float zone silicon sensors have an active thickness of $\qty{290}{\micro\meter}$ and a material budget of \mbox{$\num{3.4E{-3}} X_0$}.
Each sensor includes two rows of $\qty{50.26}{mm}$ long AC-coupled strips, in total 2032 strips.
The distance between the implants of the two rows in the center of the sensor is $\qty{68}{\micro\meter}$.
The strip pitch is $\qty{90}{\micro\meter}$, and the width of the strip implants is $\qty{22}{\micro\meter}$.
The sensors are held separated by carbon fiber reinforced aluminum bridges, which are electrically isolated against the sensor backside by Kapton strips.
For the 2S modules the nominal sensor spacing is either $\qty{1.8}{\milli\meter}$ or $\qty{4}{\milli\meter}$ depending on the position in the Outer Tracker.
The strips are connected with thin aluminum wire bonds to one of two independent front-end hybrids (FEHs). 

The silicon sensor strip signals are sampled by the CMS Binary Chips (CBCs)~\cite{Raymond:2012zz}, located on the FEHs.
One FEH comprises eight CBCs.
Each CBC collects the charge signal of $\num{254}$~strips, alternating between the strips of the top and bottom sensor to allow for the stub mechanism.
The signals from the bottom sensor are routed via a fold-over to the CBCs on the top side of the FEHs.
Each CBC channel consists of a charge sensitive amplifier followed by a threshold comparator.
The amplifier is tuned for each strip individually to compensate for different pedestal values.
The threshold value is common to all channels of a CBC.
The output signal from each channel’s comparator is then processed by the hit detect circuit to register hits in the $\qty{40}{\mega\hertz}$ digital domain.
There are four different hit detect logic modes, two of which are discussed in Section~\ref{sec:hit-detect-logic}.
The hit detect logic outputs are stored in a pipeline memory until a trigger accept signal is received.
The memory can keep events from up to $\num{512}$ consecutive bunch crossings corresponding to $\qty{12.8}{\micro\second}$.
In addition the stubs are formed on the CBC using the information from both sensors. 
The search window for the stub finding is programmable in half-strip steps up to a maximum of $\pm\num{7}$~strips.
Clusters with a configurable size limit of up to four strips are accepted for correlation.
The center position of the search window can be moved to compensate for geometric offsets that depend on the position of the CBC on the module and on the position and orientation of the module in the CMS detector.

The CBC output is compressed and transferred to the service hybrid (SEH)~\cite{Klein:2019pb} via the Concentrator Integrated Circuit (CIC)~\cite{Nodari:20192c, CMS:2022okp}. For each bunch crossing up to three stubs per CBC are forwarded to the CIC.
The data from $\num{48}$ CIC input lines (six from each CBC) are aggregated at a rate of $\qty{320}{\mega\hertz}$.
Per CBC, the lines are divided into five lines for trigger data (stubs) and a single line for data extracted when an L1 trigger accept signal is received.
In the configuration used on the 2S modules, up to 16 stubs per eight bunch crossings are transferred per CIC on its five trigger output lines through the SEH to the back-end electronics, where tracks are formed.
These tracks are used by the L1 trigger system along with information from other sub-detectors to form the L1 trigger decision.
Triggered L1 data are transferred on one data line to the SEH.

On the SEH, the low-power Gigabit transceiver {(lpGBT)}~\cite{lpGBT} serializes the data from the two CICs and passes them to the Versatile Link Plus transceiver (VTRx+)~\cite{TROSKA2023168208}, which sends the data to the back-end electronics after opto-electrical conversion.
Two DC-DC converters are used to convert a nominal input voltage of between $\qty{5.5}{V}$ and $\qty{12}{V}$ to the required low voltages of $\qty{1.25}{V}$ and $\qty{2.5}{V}$~\cite{dcdc:2023}.

\begin{figure}[!tb]
  \centering
  \includegraphics[width=\textwidth]{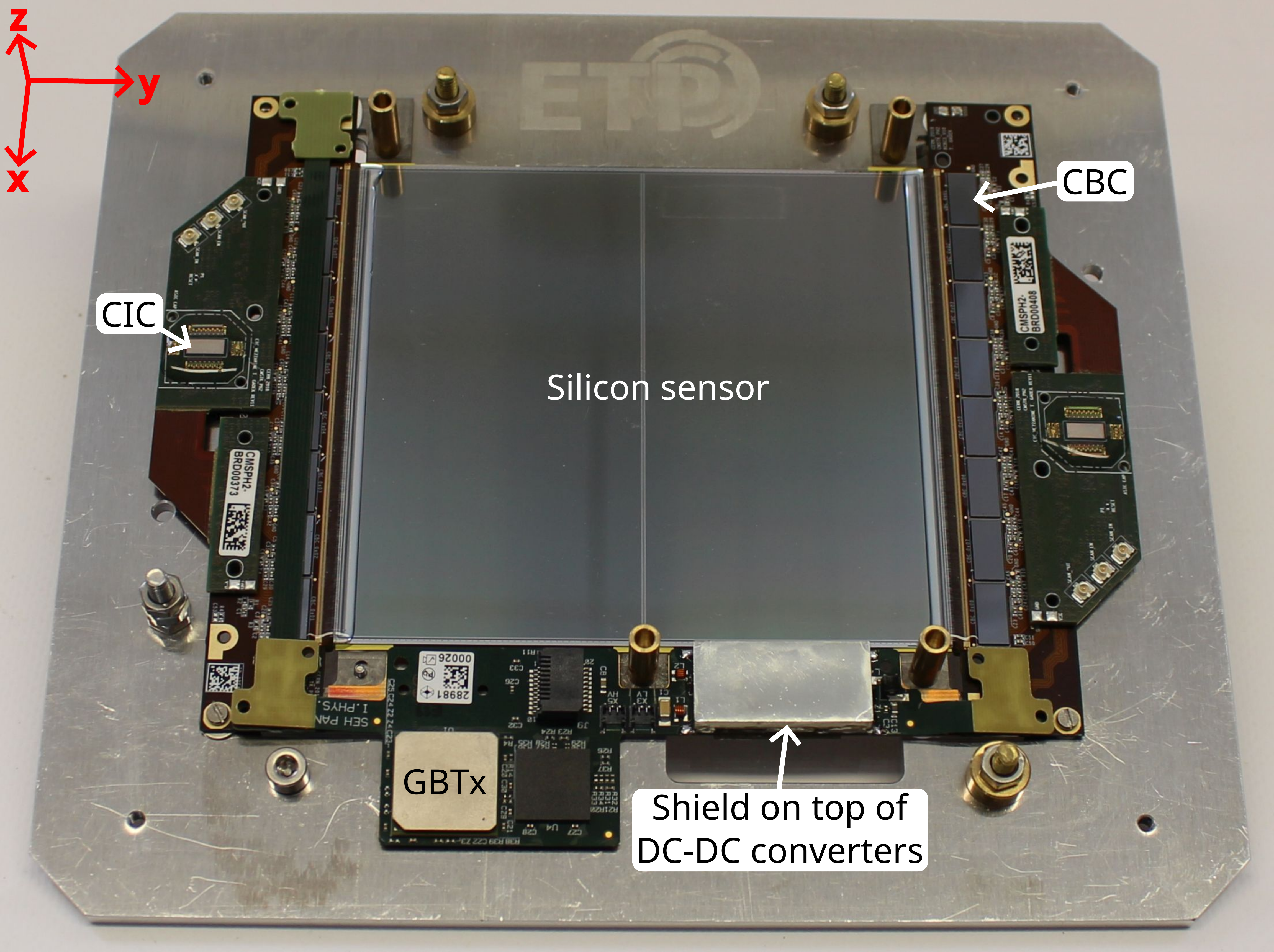}
  \caption[Beam test {2S} module]{Photo of a {2S} prototype module used in the beam test campaigns.
  The module is screwed onto its aluminum module carrier which is then connected to a larger support structure to position the module in the beam test setup.
  The {VTRx} module is not plugged in this picture.
  The local coordinate system for the 2S module is indicated on the top left corner of the photo.
  \label{fig:2s-beamtest-module}}
\end{figure}

In the beam test campaigns presented, the components of the {2S~modules} differed from those described above.
One prototype module is shown in Figure~\ref{fig:2s-beamtest-module}.
CBCs of the final version 3.1 were used on prototype FEHs without CICs.
The CIC ASICs in version 1 or version 2 were mounted on separate printed circuit boards, which were connected to the FEH via two fine-pitch connectors on the top side of the hybrid. 
Only the right-hand side version of the FEH was available.
To read out the left side of the module, a special flat Kapton flex cable was used to connect the SEH version~3.1 to a right-hand side FEH.
Instead of the lpGBT, its predecessor {GBTx}~\cite{Moreira:2009pem} was used together with the {VTRx}~\cite{TROSKA2023168208} instead of the {VTRx+}.
For the DC-DC conversion on the SEH, the predecessor chip FEAST~2.1~\cite{dcdc:2023} and a commercial buck converter were used instead of the final ASICs. 
An input voltage of $\qty{10.5}{\volt}$ was applied. 
The sensor spacing of the tested modules was measured to be about $\qty{1.65}{\milli\meter}$ instead of the nominal $\qty{1.8}{\milli\meter}$.
In the beam test campaigns seven 2S prototype modules were installed.
Table~\ref{tab:all_modules} contains the list of the five prototype modules that have been extensively tested and were used for the analyses presented in this paper. The influence of the usage of prototype components on the results is negligible. 

\begin{table}[!tbp]
    \centering
    \caption{List of the five 2S prototype modules used in the data analysis of the beam test campaigns. All modules have a nominal sensor spacing of $\qty{1.8}{mm}$ and use the VTRx and GBTx for readout. All modules use CBC version~3.1 and SEH version~3.1.}
   \label{tab:all_modules}
    \vspace{0.2cm }
    \begin{tabular}{cccc}
    \hline
    Module name & Fluence &  CIC      & Beam test date    \\ \hline
    module 1    & 0 & version $1$ & 11/2019 \\ 
    module 2    & 0 & version $1$ & 11/2019\\ 
    module 3    & 0 & version $1$ & 11/2019 \\ 
    module 4    & $\num{4.6E14}\,\textrm{n}_{\textrm{eq}}\textrm{cm}^{-2}$ & version $2$ & 08/2020 \\ 
    module 5    & 0 & version $2$ & 11/2020 \\ 
    \hline
    \end{tabular}
\end{table}

%% file: DN-20-014_Setup-and-DAQ.tex
\section{Beam Test Setup, Data Acquisition, and Track Reconstruction} \label{sec:beamtestsetup}

\begin{figure}[!bt]
    \centering
    \includegraphics[width=\textwidth]{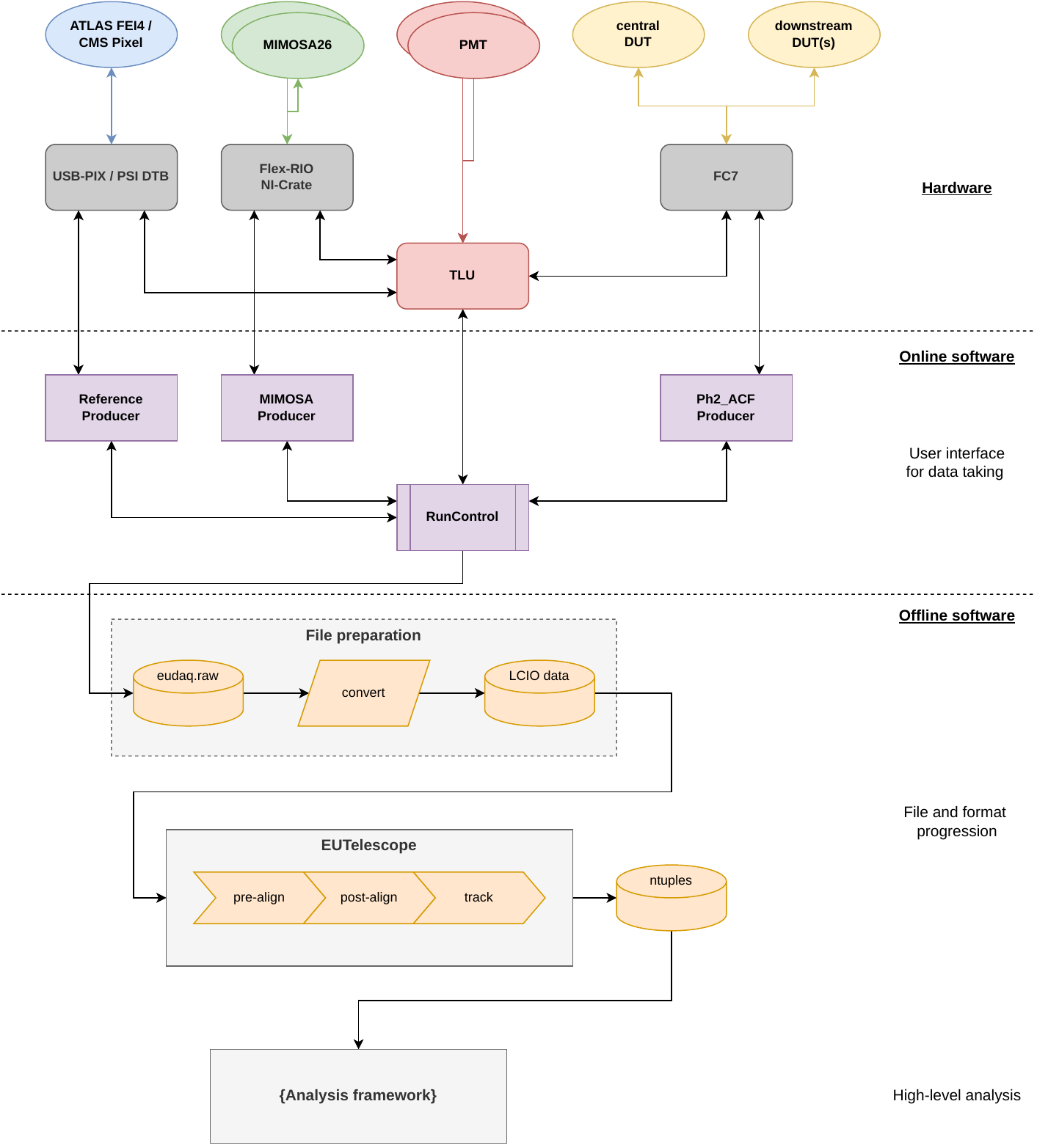}
    \caption[]{The data acquisition (DAQ) scheme used for data taking. It is split into hardware components and online and offline software packages.}
    \label{fig:daqChain}
\end{figure}

The {DESY-II} Test Beam Facility provides an electron beam with electron momenta of up to $\qty{6.3}{GeV}$.
The size of the beam spot is adjustable with a lead collimator.
The size chosen was $\qtyproduct{12 x 20}{\milli\meter}$ to match the telescope acceptance. 
The hardware and software components used for data acquisition are summarized in the data flow scheme shown in Figure~\ref{fig:daqChain}.
The individual objects are introduced below.

\begin{figure}[!tb]\centering
    \includegraphics[width=0.99\textwidth]{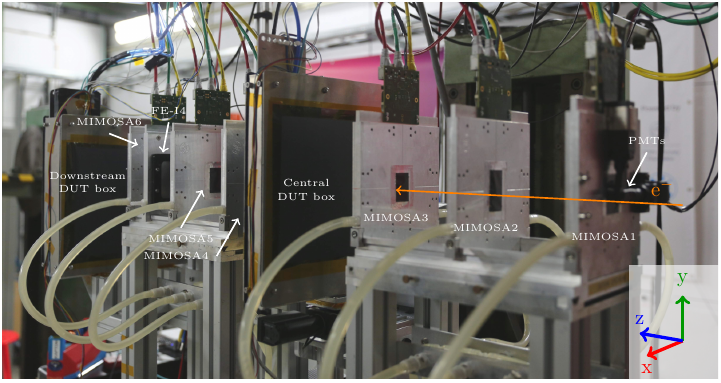}
    \caption[{DESY} infrastructure with {2S} modules]{Picture of the beam test setup. The central {DUT} box is installed between the three upstream (MIMOSA1--3) and three downstream (MIMOSA4--6) telescope planes. Three additional modules were installed in the downstream {DUT} box. The timing reference detector (FE-I4) is installed between telescope planes 5 and 6.}
    \label{fig:dutsTelescope}
\end{figure}

\subsection{Hardware Components} \label{sec:setup-hardware-components}

The {EUDET}-type beam telescopes~\cite{Jansen:2016bkd} available at the facility were used during the beam tests.
One beam test setup is shown in Figure~\ref{fig:dutsTelescope}.
The telescope consists of six {MIMOSA26} active pixel devices with an active area of approximately $\qtyproduct{1 x 2}{\centi\meter}$ read out with the {EUDAQ} framework~\cite{Liu:2019wim}.
The material budget in units of radiation lengths $X_0$ of a {MIMOSA26} plane including a protective Kapton foil on each side is \mbox{$\num{7.5E{-4}} X_0$} assuming a sensor thickness of $\qty{54}{\micron}$~\cite{Jansen:2016bkd}.
The small pixel size of $\qtyproduct{18.4 x 18.4}{\micron}$ provides a very good pointing resolution of about $\qty{10}{\micron}$ for the track impact position on the device under test (DUT).
The telescope planes have a long integration time of more than $\qty{100}{\micro\second}$ and accumulate multiple tracks within an event, only one of which can trigger the readout of the {2S~modules}.
Because of this, a timing layer is inserted between the fifth and sixth telescope layer to provide timing information at a readout frequency of $\qty{40}{\mega\hertz}$.
This timing layer is called reference detector in the following. 
Depending on the specific beam test campaign, either an ATLAS FE-I4 pixel module~\cite{Obermann:2014goa} or a CMS Phase-1 pixel module~\cite{PhaseIPixelUpgrade} was used.

The readout of all detector devices is organized and triggered by an {EUDET}-type Trigger Logic Unit ({TLU})~\cite{tlu}, also synchronizing event numbering across the different detector devices.
The TLU uses the coincidence of two crossed scintillator planes placed in front of the telescope to generate the trigger.
The timing resolution of the scintillators was measured to be $\qty{1}{\nano\second}$ or better~\cite{Kroger:2021wvs}.
The primary detector under test was located in the central DUT box (Figure~\ref{fig:dutsTelescope}) between the three upstream and three downstream telescope planes, to benefit from the best possible track pointing resolution.
The central DUT box was mounted on a movable table, which allows positional scans in the sensor plane of the module as well as rotational scans around an axis.
Rotation around the silicon strip axis is used to emulate particles with different \pt.
The bottom sensor was facing the upstream arm of the telescope. 
Additional {2S~modules} were mounted in a fixed position in a downstream DUT box behind the last plane of the beam telescope for synchronization studies and to test simultaneous optical readout.
The orientation of the modules was the same as that of the primary {DUT}.
All {DUT} boxes were light-tight and purged with nitrogen gas to ensure low humidity around the sensors and to cool the electronic components through the gas flow.
Temperature and humidity in both boxes were monitored.
In addition, the leakage current of each module was constantly monitored during the beam tests.
The {2S~module} with the irradiated sensors was cooled down to sensor temperatures of about $\qty{-17}{\celsius}$ for the measurements.
The module was mounted on a copper jig through which ethanol circulated, cooled by an external cooler. 
The modules with non-irradiated sensors were operated at about $\qty{+23}{\celsius}$ room temperature.

All {2S~modules} were connected to the {FC7} FPGA readout board~\cite{Pesaresi_2015} via optical fibers.
The {FC7} is a $\upmu$TCA standard compatible Advanced Mezzanine Card built around the Xilinx Kintex-7 FPGA, which is capable of supporting line rates up to $\qty[per-mode = symbol]{10}{\giga\bit\per\second}$.
It provides two FPGA Mezzanine Card (FMC) sockets, enabling various configurable I/O add-ons.
One FMC is used as the interface to the optical fibers of the 2S~modules. The other socket holds the DIO5 card~\cite{DIO5} equipped with LEMO connectors to receive the trigger signal from the TLU.

A global coordinate system is used for the telescope.
Its origin lies in the center of the first telescope plane. 
The $z$-axis points downstream along the beam direction, the $y$-axis points upwards, and the $x$-axis points in the horizontal plane.
The resulting right-handed coordinate system is shown in Figure~\ref{fig:dutsTelescope}.
In addition, local coordinate systems are defined for the {2S~modules}.
For each module, the origin is at the center of the bottom sensor.
The $z$-axis points up through the module, the $y$-axis points along the strips and the $x$-axis points along the front-end hybrids perpendicular to the strips. The geometric center of the sensor is at $x=y=\qty{0}{\milli\meter}$. At $y=\qty{0}{\milli\meter}$ the strips of the left side meet with those of the right side on the sensor.
The resulting right-handed coordinate system is illustrated in Figure~\ref{fig:2s-beamtest-module}.

\subsection{Data Flow and Software Packages} \label{sec:setup-flow-and-sw}
In the telescope's {EUDAQ} software framework~\cite{Liu:2019wim}, each detector system is managed by its own sub-process, called producer, which is linked to the main run control program.
The producer for the telescope modules is the so-called MIMOSA producer.
The {2S~modules} are handled by the data streaming module of the Phase-2 Acquisition and Control framework ({Ph2\_ACF}).
The run control provides a state machine that coordinates all measurements, aggregates the data and writes the collection to an event-based raw data file.
In preparation for further analysis, the data files are converted to the Linear Collider I/O ({LCIO})~\cite{Gaede:2003ip} data format.

In the {EUTelescope} software framework~\cite{Bisanz_2020}, the beam test data are used to reconstruct particle tracks using information from the telescope planes.
After identifying the barycenter of clusters of neighboring hit pixels or strips for each individual detector plane, the telescope information is used for track fitting and alignment using the {General Broken Lines} approach~\cite{KLEINWORT2012107}.
In the telescope track fit, the material budget of the DUTs and the reference plane are taken into account.
The position alignment of the DUTs and the reference plane is performed based on the track information by minimizing the residuals between the track impact points and the reconstructed hits.
Alignment is performed for the two translations perpendicular to the beam axis and all three rotation angles.
Each sensor plane of the {2S~modules} is aligned independently, providing information about the module's mounting accuracy.
As a final step, all information about cluster positions and track intersection points for each detector layer are stored for further analysis.

%% file: DN-20-014-Reconstruction_dataanalysis.tex
\section{Data Analysis Definitions} \label{sec:analysis}
    The following definitions are used in the analysis:

    \paragraph*{Bias voltage}
    The {2S} sensors are biased by applying a negative potential to the backplane contact on the unstructured sensor backside and the ground potential to the bias ring on the structured sensor frontside. 
    In the following, only absolute values are quoted for the bias voltage levels.
    Unless stated otherwise, data were collected at a bias voltage $\mathrm{U} = \qty{300}{V}$ for modules with unirradiated sensors and $\qty{600}{V}$ for the module with irradiated sensors.
    {2S~modules} are designed to operate at bias voltages up to $\qty{800}{V}$.
    
    \paragraph*{Noise occupancy} 
    The noise occupancy of {2S~modules} is measured with the electron beam off, using random triggers generated internally on the {FC7} {FPGA} board. The noise occupancy is defined as the probability to detect one noise hit per readout channel and event.

     \paragraph*{Thresholds}
     During the first data-taking runs for each beam test, a suitable threshold was identified by monitoring the module hit maps while scanning several threshold values.
     A threshold of approximately $\qty{6000}{electrons}\,(\textrm{e}^{-})$ was chosen for data taking for both the irradiated and unirradiated modules.
     The threshold value is configured via the firmware as a parameter.
     Conversion to electrons uses the modules' measured pedestal values and an assumed common conversion factor \cite{LastBTPaper}.
     The influence of the threshold setting on the data taking is discussed in Section~\ref{subsec:SignalNoise}.

    \paragraph*{Track selection} \label{sec:analysis:track_selection} 
    Only tracks with hits in all {MIMOSA26} planes are taken into account.
    To accommodate the limited time resolution of the {MIMOSA26} planes, it is necessary that the track projection on the reference plane (ATLAS FE-I4 or CMS pixel module) is linked to a reference plane cluster with the distance constraints $\Delta x_{\textrm{ref}}(\textrm{track},\textrm{cluster})\le \qty{0.2}{mm}$ and $\Delta y_{\textrm{ref}}(\textrm{track},\textrm{cluster})\le \qty{0.08}{mm}$, when using the ATLAS FE-I4.
    When using the CMS pixel module the constraints are $\Delta x_{\textrm{ref}}(\textrm{track},\textrm{cluster})\le \qty{0.15}{mm}$ and $\Delta y_{\textrm{ref}}(\textrm{track},\textrm{cluster})\le \qty{0.1}{mm}$.
    
    \paragraph*{Track isolation}
    If there is no additional track projection onto the reference plane within a conservatively chosen radius of $\qty{0.6}{mm}$, the track is called isolated.
    The number of isolated tracks is denoted as $n_{\textrm{isolated tracks}}$.

\begin{figure}[!tbp]
      \centering
      \includegraphics[width=0.72\textwidth]{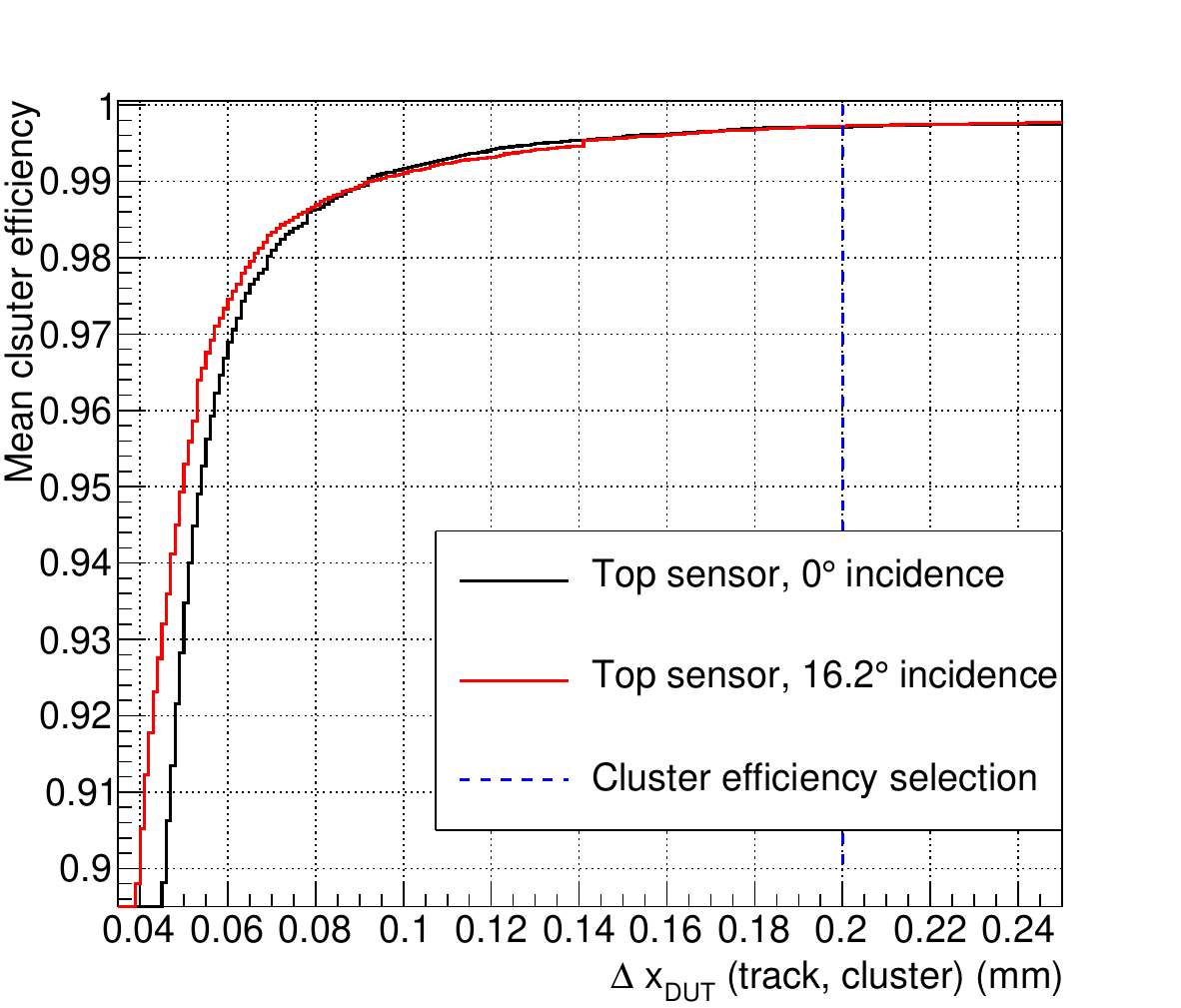}
      \caption[]{Mean cluster efficiency of the top sensor of module~1 at perpendicular incidence (black line) and at $\qty{16.2}{\degree}$ incidence angle (red line) as a function of the cluster efficiency distance condition $\Delta x_{\textrm{DUT}}(\textrm{track},\textrm{cluster})$. The blue dashed line marks the chosen condition value.}
      \label{fig:htdCutScan}
    \end{figure}

    \paragraph*{Cluster efficiency} \label{sec:analysis:efficiencies}
    By comparing the projected intersection of isolated tracks with the sensors of the central DUT, the number of tracks linked to the DUT that satisfy the distance criteria $\Delta x_{\textrm{DUT}}(\textrm{track},\textrm{cluster}) \le \qty{0.2}{mm}$ is defined as $n_{\textrm{DUT linked tracks}}$. 
    The cluster efficiency $\varepsilon_{\textrm{cluster}}$ is defined as
    \begin{equation}
      \varepsilon_{\textrm{cluster}} = \frac{n_{\textrm{DUT linked tracks}}}{n_{\textrm{isolated tracks}}}~.
      \label{eq:cluster-efficiency-definiton}
    \end{equation}
    The cluster efficiency is evaluated independently for the two sensors in the {2S~module}.
    Figure~\ref{fig:htdCutScan} shows the mean cluster efficiency of the top sensor of module~1 as a function of the cluster efficiency distance condition $\Delta x_{\textrm{DUT}}(\textrm{track},\textrm{cluster})$ at perpendicular incidence and at an angle of $\qty{16.2}{\degree}$.
    At the chosen value of $\Delta x_{\textrm{DUT}}(\textrm{track},\textrm{cluster})=\qty{0.2}{\milli\meter}$ the efficiency at both incidence angles is constant and similar.
    As the beam spot is much smaller than the size of the active area of the DUTs and centered on the module (if not stated otherwise), no fiducial region has to be defined.

    \paragraph*{Stub efficiency}
    Two definitions are used for the stub efficiency.
    As with the cluster efficiency, it is possible to compare the projected intersection of isolated tracks on both 2S sensors with the stub information sent by the module.
    The number of tracks fulfilling the criteria 
    \begin{displaymath}
        \Delta x(\textrm{track},\textrm{stub position})\le \qty{0.2}{mm}
    \end{displaymath}  for the seed layer and \begin{displaymath}
        \Delta x(\textrm{track},\textrm{stub position + stub bend})\le \qty{0.2}{mm}
    \end{displaymath} for the correlation layer is defined as $n_{\textrm{stub linked tracks}}$.
    The stub position is defined as the cluster position in $x$ on the seed layer.
    The stub bend is defined as the displacement in $x$ between the cluster on the correlation layer and the center of the search window.
    The module stub efficiency $\varepsilon_{\textrm{stub}}^{\textrm{mod}}$ is given by
    \begin{equation}
      \varepsilon_{\textrm{stub}}^{\textrm{mod}} = \frac{n_{\textrm{stub linked tracks}}}{n_{\textrm{isolated tracks}}}~.
    \end{equation}
    This definition includes all possible sources of inefficiency of the module to form a stub, but most importantly the hit efficiency of the two individual sensor layers.
    
    To characterize the functionality of the CBC stub logic alone without the influence of hit detection inefficiencies, the CBC stub efficiency $\varepsilon_{\textrm{stub}}^{\textrm{CBC}}$ is introduced.
    Only events that have clusters fulfilling the cluster efficiency criteria in both the bottom and top sensors and in which each cluster is assigned to the same track are considered.
    In addition, cluster widths in both sensors are limited four or fewer strips, since the on-chip stub detection logic rejects clusters wider than four strips.
    The number of tracks fulfilling these criteria is defined as $n_{\textrm{offline stubs}}$.
    Counting the number of stubs $n_{\textrm{matched stubs}}$ fulfilling the criteria $\Delta x(\textrm{stub position},\textrm{cluster center})\le \qty{1}{strip}$ for the seed layer and $\Delta x(\textrm{stub position + stub bend},\textrm{cluster center})\le \qty{1}{strip}$ for the correlation layer the CBC stub efficiency is defined as
    \begin{equation}
      \varepsilon_{\textrm{stub}}^{\textrm{CBC}} = \frac{n_{\textrm{matched stubs}}}{n_{\textrm{offline stubs}}}~.
    \end{equation}
\paragraph*{Efficiency uncertainties}
    All efficiencies are shown with statistical uncertainties only. Systematic uncertainties are  studied in Ref.~\cite{PhdTimZiemons} using module~1 and are estimated to be in the order of $\qty{0.06}{\percent}$ for cluster efficiencies 
    at perpendicular incidence.
    This estimation is based on the variation of the various event selection criteria and is dominated by the selection of the TDC phase (details in section \ref{sec:asynchronous_readout}) and the choice of $\Delta x_{\textrm{DUT}}(\textrm{track},\textrm{cluster})$.
    At an incidence angle of $\qty{16.2}{\degree}$, which is close to the expected \pt{} discrimination threshold at an acceptance window of $\pm \qty{5}{strips}$, the systematic uncertainties increase to about $\qty{0.12}{\percent}$.
    The values are absolute efficiency uncertainties.

  \subsection{Hit Detection Modes} \label{sec:hit-detect-logic}
  
\begin{figure}[!tbp]
      \centering
      \includegraphics[width=0.72\textwidth]{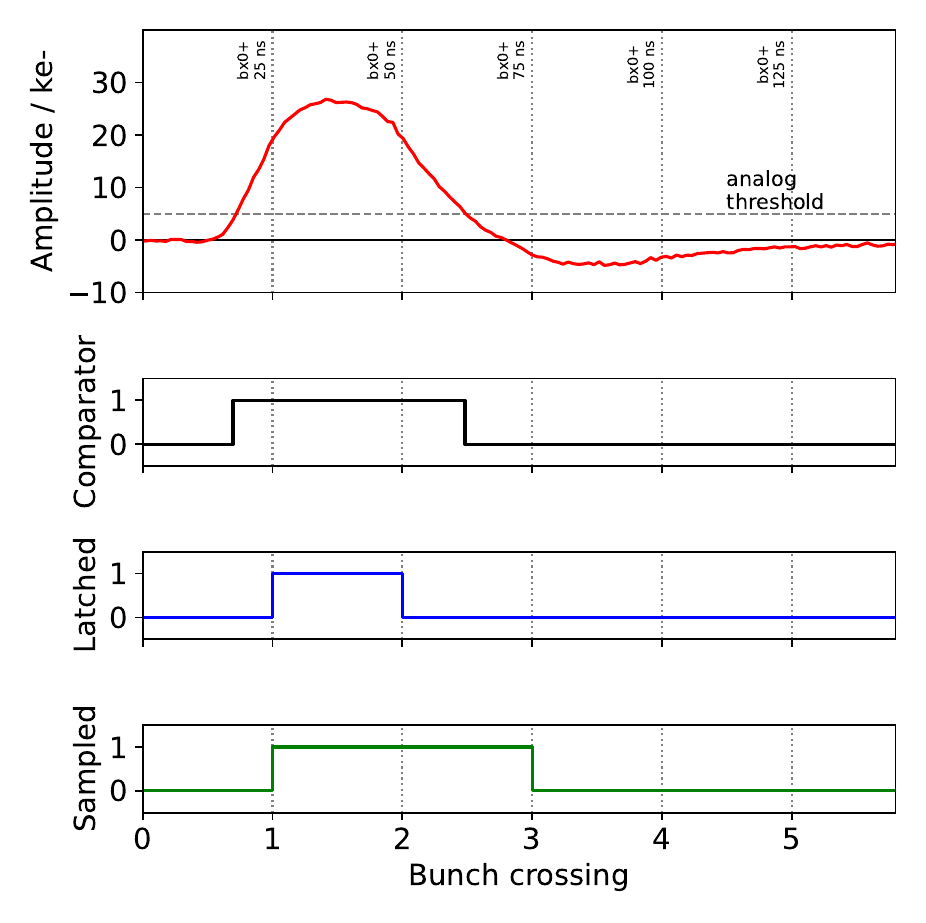}
      \caption[]{Comparison of the different hit detection modes. The top panel shows the analog pulse reconstructed from a single charge-injected pulse. The pulse height is approximately $\num{25000}~\textrm{e}^{-}$. Applying a threshold leads to a digital comparator output signal, which is converted differently by the latched and the sampled hit detection modes.
      }
      \label{fig:hit-detect-logic-illustration}
    \end{figure}

    The {CBC} implements four different hit detection logic modes~\cite{CBC3_1}, of which two were studied during these beam test campaigns: fixed-pulse-width and $\qty{40}{MHz}$ sampled-output, referred to as latched mode and sampled mode, respectively.
    In latched mode, the high comparator output is captured at any given time and the hit is reported for the following clock cycle only.
    A subsequent hit will only be sent out after the signal returns below threshold.
    In sampled mode, the channel's signal comparator output is captured on the rising $\qty{40}{MHz}$ clock edge and a hit is reported as long as the comparator output is high when sampled.
    Figure~\ref{fig:hit-detect-logic-illustration} shows the output of these two modes on an example signal pulse as a function of time.
    The two modes have different benefits depending on the detector occupancy.
    In these studies, no difference in the detector efficiency is expected, due to the low rate of the electron beam. Latched mode is intended for operation at the HL-LHC.
    
  \subsection{Asynchronous readout}\label{sec:asynchronous_readout}
    Since the electron beam is asynchronous to the $\qty{40}{MHz}$ readout clock of the DUT, particles and therefore the trigger signal arrive at an arbitrary clock phase.
    Due to the fast shaper response in comparison to the clock frequency and depending on the arrival time of the particle, the sampling time of the signal is not always at the maximum of the pulse amplitude (Figure~\ref{fig:hit-detect-logic-illustration}).
  
\begin{figure}[!tbp]
      \centering
      \includegraphics[width=0.8\textwidth]{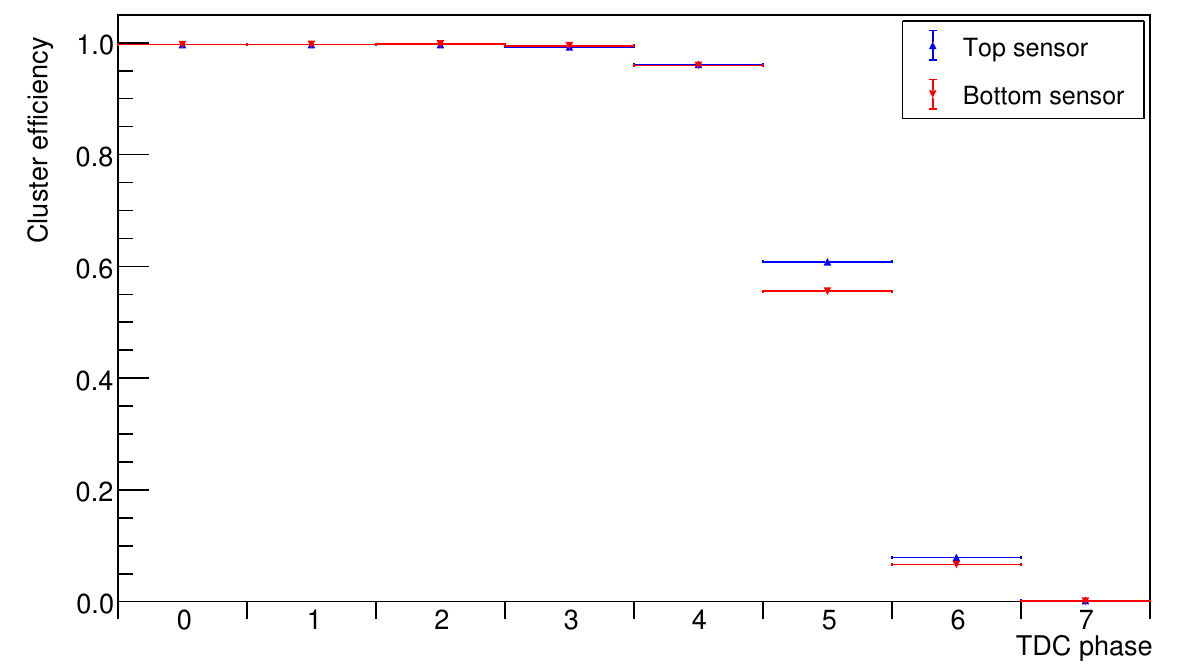}
      \caption[]{
        Distribution of cluster efficiencies as a function of the TDC phase measured in sampled mode.
        In this example TDC phase bins 0 and 1 are selected for the analysis. One bin corresponds to a \qty{3.125}{ns} delay.
      }
      \label{fig:tdc-cluster-efficiency}
    \end{figure}

    This signal sampling at a lower pulse height can lead to a reduction of the detection efficiency.
    When operated at the LHC, the modules will be synchronized to the accelerator clock and the sampling time can be fixed at the peak of the signal pulse.
    To compensate for this effect at the beam test, the {FC7} firmware measures the time of arrival of each trigger signal from the {TLU} with a granularity of $\qty{3.125}{ns}$ with respect to the $\qty{40}{MHz}$~clock.
    This quantity can thus take eight values and is referred to as TDC phase.
    Events with the optimal phase can be selected for the analysis, and events with off-peak sampling are rejected.
    Figure~\ref{fig:tdc-cluster-efficiency} depicts one example of the reconstructed cluster efficiency as a function of the TDC phase for both sensors of a {2S~module}. 
    The two most efficient TDC phases are selected for each run individually in all further analyses. 

%% file: DN-20-014_results.tex
\section{Results and Discussion} \label{sec:results}

All presented analyses are done for the central DUT and at perpendicular beam incidence if not stated otherwise.

\subsection{Tracking and Module Resolution}

The alignment of the {2S~modules} within the beam telescope is optimized by iteratively minimizing the absolute value of the residuals between the track and cluster coordinates in $x$ (perpendicular to the strips).
This test provides information about the resolution of the {2S~module} and the uncertainty of the telescope prediction at the {DUT} plane.

\subsubsection*{Tracking resolution:}

Since the sensors on the {2S~module} have a strip pitch of $\qty{90}{\micro\meter}$ and the readout is binary, the expected deviation between the physical and the detected hit position for events with clusters of only a single strip is modeled as a rectangular function of width \mbox{$w = \qty{90}{\micro\meter}$}.
To account for a finite telescope resolution, the rectangular function is convolved with a Gaussian distribution centered around zero with a standard deviation $\sigma_\textrm{t}$ reflecting the uncertainty of the track impact position on the DUT.
The calculation of the convolution results in the subtraction of two error functions:

\begin{align}
    P_\textrm{res}(x) &= \int_{-\infty}^{\infty} \left[ \Theta \left( t+\dfrac{w}{2} \right) - \Theta \left( t-\dfrac{w}{2} \right) \right] \cdot \dfrac{1}{\sqrt{2 \pi \sigma_\textrm{t}^2}} \exp{ \left( - \dfrac{(t-x)^2}{2 \sigma_\textrm{t}^2} \right)} \textrm{d}t \nonumber \\
      &= \dfrac{1}{2} \left[ \text{erf}\left( \dfrac{x+\dfrac{w}{2}}{\sqrt{2\sigma_\textrm{t}^2}}\right)-\text{erf}\left( \dfrac{x-\dfrac{w}{2}}{\sqrt{2\sigma_\textrm{t}^2}}\right) \right]
      \label{eq:convModelResolution}
\end{align}
where $\Theta$ is the Heaviside step function.

The fit to the residual distribution is performed with the width and a shift of the rectangle function as well as the standard deviation $\sigma_\textrm{t}$ of the Gauss distribution and a scaling factor as fit parameters.
The fit is applied only in the definition range of the model, $\qty{-45}{\micro\meter} < x - \text{shift} < \qty{45}{\micro\meter}$.
Only clusters with a cluster size in $x$ of one are accepted for this study.
Applying this selection, the distribution of hit detection coordinates within a single strip is non-uniform as the probability of larger clusters is increased in the border region between two strips.
To cope with this effect and to better represent the model, the distribution of the residuals is normalized by the number of measured single strip clusters as a function of the track position relative to the nearest strip center~\cite{PhdTimZiemons}.

\begin{figure}[b]\centering
    \includegraphics[width=0.6\textwidth]{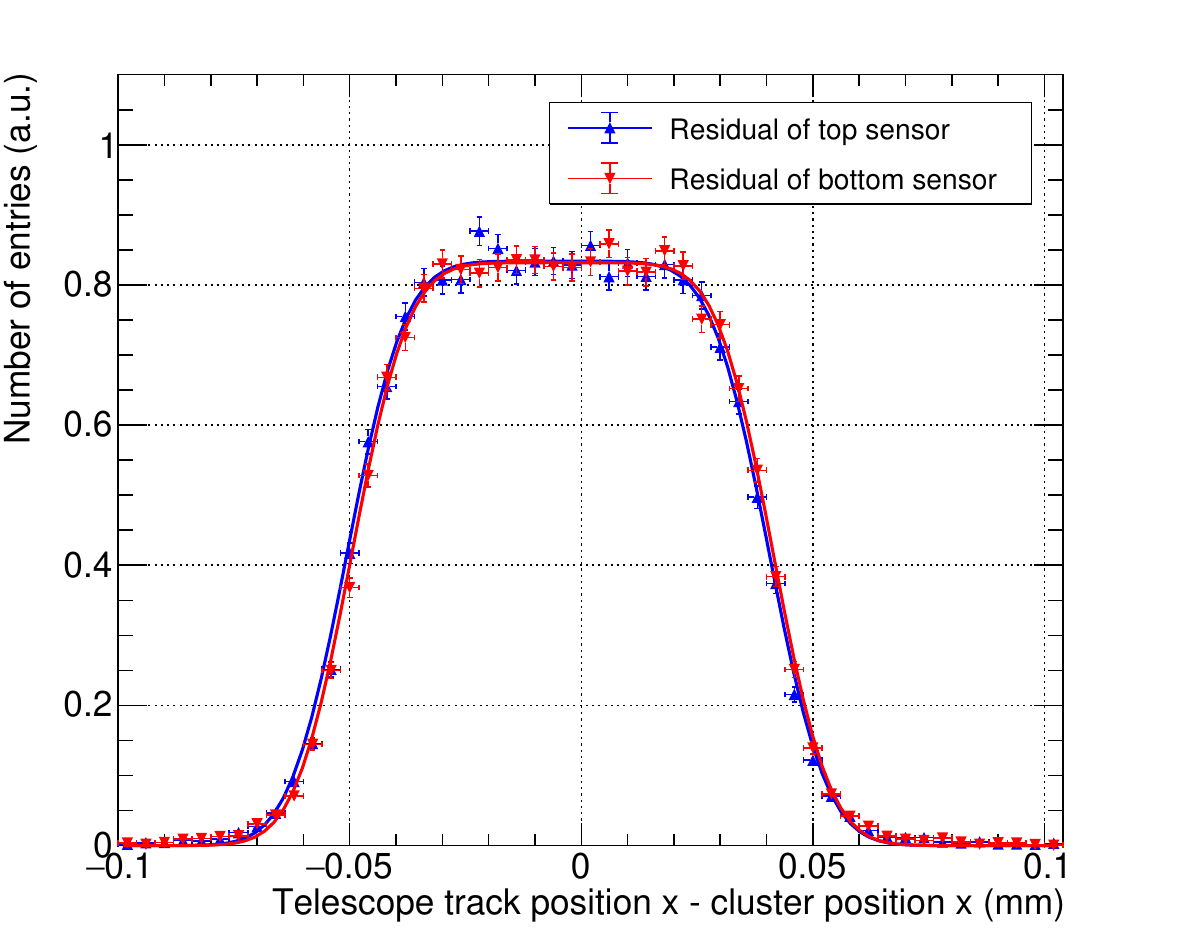}
    \caption[]{The residual distributions for the two sensors of the unirradiated module~1. The residual distributions of the top sensor (blue markers) and bottom sensor (red markers) are shown together with the fits of the convolution model according to Equation~\ref{eq:convModelResolution} (colored lines). 
    }
    \label{fig:resolutionACM1804}
\end{figure}

Figure~\ref{fig:resolutionACM1804} shows the residual distribution for both sensors of the unirradiated module~1 at a beam energy of $\qty{4}{\giga\electronvolt}$.
The fit parameters are summarized in Table~\ref{tab:ResolutionFitAcM1804}.
The fit results of the left and right rectangle borders are in good agreement with the expectation of \mbox{$w = \qty{90}{\micro\meter}$}.
The fitted values of the rectangle shift of \mbox{$\left(-4.9\pm0.3\right)\unit{\micron}$} and \mbox{$\left(-4.1\pm0.3\right)\unit{\micron}$} of the top and bottom sensor reflect the alignment precision.
The uncertainty of the track impact position is calculated to be \mbox{$\left({9.4}_{-0.4}^{+0.3}\right)\unit{\micron}$} using the GBL Track Resolution Calculator~\cite{spannagel_2016_48795}.
The respective uncertainties result from varying the beam energy of $\qty{4}{\giga\electronvolt}$ by $\qty{5}{\percent}$ to account for an energy spread within the beam telescope.
The fit result of the track uncertainty agrees with the calculation.

\begin{table}[tb]
    \centering
    \caption{Summary of parameters of the fits to the residual distributions shown in Figure~\ref{fig:resolutionACM1804}.}
    \label{tab:ResolutionFitAcM1804}
    \vspace{0.2cm}
    \begin{tabular}{cccc}
    \hline
    Parameter & Top sensor & Bottom sensor & Expectation \\
    \hline
    $\chi^2$/ndf & $13.72$ / $18$ & $10.86$ / $18$ & $-$\\
    Shift ($\qty{}{\micro\meter}$) & $-\,4.9\pm0.3$& $-\,4.1\pm0.3$  & $0$ \\
    Width ($\qty{}{\micro\meter}$) & $91.3\pm1.1$  & $91.1\pm1.1$  & $90$ \\
    $\sigma_\textrm{t}$ ($\qty{}{\micro\meter}$) & $10.0\pm0.9$  & $9.7\pm0.8$ & ${9.4}\,{}_{-\,0.4}^{+\,0.3}$ \\
    \hline
    \end{tabular}
\end{table}

\subsubsection*{Module spatial resolution:}

The standard deviation of the residual distribution without the cluster size selection applied is $\qty{28.7}{\micro\meter}$ for the top sensor and $\qty{28.8}{\micro\meter}$ for the bottom sensor.
The intrinsic DUT single point resolution $\sigma_{\text{DUT}}$ can be calculated by quadratically subtracting the telescope uncertainty from these values.
The resolution of the sensors is calculated to be $\sigma_{\text{DUT,top}} = \left(26.9\pm0.3\right)\unit{\micron}$ for the top sensor and $\sigma_{\text{DUT,bottom}} = \left(27.1\pm0.3\right)\unit{\micron}$ for the bottom sensor.
The expected resolution of a detector with binary readout and a sensor strip pitch of $\qty{90}{\micro\meter}$ is $\sigma_\text{expect} = \qty{26.0}{\micro\meter}$, assuming a flat detection probability distribution.
The result is in acceptable agreement with the expectation.

\subsection{Signal and Noise} \label{subsec:SignalNoise}

\begin{figure}[!tbp]\centering
\includegraphics[width=\textwidth]{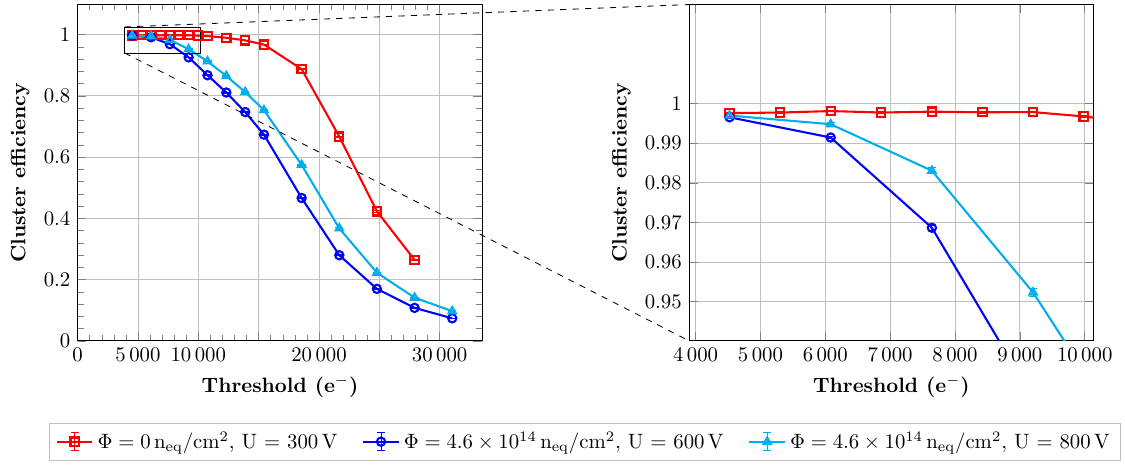}
\caption[]{The cluster efficiency as a function of the threshold of module~5 (unirradiated) and module~4 assembled with sensors irradiated with protons to a fluence of $4.6 \times 10^{14}\,\textrm{n}_{\textrm{eq}}\textrm{cm}^{-2}$.
The red data points indicate cluster efficiencies measured with the unirradiated module at a bias voltage of $\qty{300}{\volt}$.
Data from the irradiated module are shown in blue for a bias voltage of $\qty{600}{\volt}$ and in cyan for a bias voltage of $\qty{800}{\volt}$.
The right plot shows a zoomed version of the left plot.
Only data from the bottom 2S sensors are shown.
Statistical uncertainties are shown as error bars.
However, uncertainties are dominated by systematics, estimated in Section~\ref{sec:analysis}. 
}
\label{fig:KIT-Thresholdscan-Comparison-Latched}
\end{figure}

To investigate the influence of the threshold on the signal and noise behavior of 2S modules, measurements were made with different threshold settings, called threshold scans.
A comparison of the threshold scans for an unirradiated module and the module assembled with irradiated sensors is shown in Figure~\ref{fig:KIT-Thresholdscan-Comparison-Latched}. 
The unirradiated sensors, operated at a bias voltage of $\qty{300}{V}$, reach cluster efficiencies of $\qty{99.7}{\percent}$ up to a threshold of $\num{10000}\,\textrm{e}^{-}$.
The irradiated sensors deliver smaller signals due to radiation damage. 
Thus, the maximum cluster efficiency measured at the bias voltage of $\qty{600}{V}$ and thresholds between $4000\,\textrm{e}^{-}$ and $6000\,\textrm{e}^{-}$ is slightly lower compared to the unirradiated module.
By increasing the bias voltage to $\qty{800}{V}$, it is possible to increase the cluster efficiency of the irradiated module~4 to $\qty{99.5}{\percent}$ at a threshold of $6000\,\textrm{e}^{-}$.
  
\begin{figure}[!tbp]\centering
    \includegraphics[width=.6\textwidth]{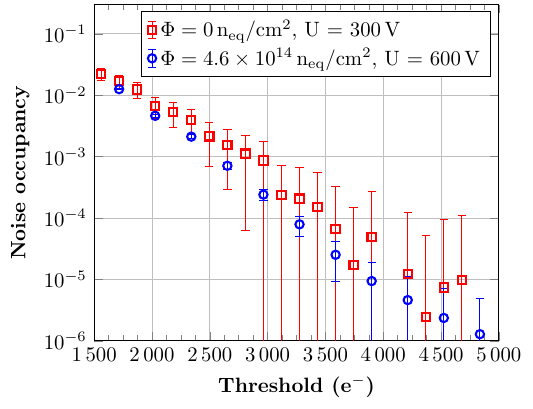}
    \caption[]{Noise occupancy as a function of the threshold for module~5 (unirradiated, red markers) and module~4 (irradiated sensors, blue markers).
    Data from the unirradiated module are taken at a bias voltage of $\qty{300}{V}$ and at room temperature.
    Data from the irradiated module are measured at a bias voltage of $\qty{600}{V}$ and at a sensor temperature of approximately $-17^\circ\textrm{C}$.
    The number of events per threshold setting is significantly larger for the measurement with the irradiated module.
    The error bars indicate statistical uncertainties. 
    }
\label{fig:KIT-Thresholdscan-Comparison-Latched-Noise}
\end{figure}

For comparison, the noise occupancy level is plotted as a function of the threshold in Figure~\ref{fig:KIT-Thresholdscan-Comparison-Latched-Noise}.
Because the measurements with the unirradiated module have been performed at room temperature, the noise level is larger compared to the measurement with the irradiated module, during which the sensors were cooled to $\qty{-17}{\celsius}$.
The noise measurement for the unirradiated module was performed with a smaller number of events.
For thresholds above $\num{4500}\,\textrm{e}^{-}$ the noise occupancy is smaller than $\num{E-5}$ for both the unirradiated and irradiated sensors.
Channel occupancy during HL-LHC operation is expected to be up to about one percent in the CMS Outer Tracker.
The noise occupancy is therefore more than three orders of magnitude lower than the expected signal occupancy.
At a bias voltage of $\qty{800}{V}$, the noise occupancy of the irradiated module coincides with the distribution at $\qty{600}{V}$ and, therefore, is not shown in Figure~\ref{fig:KIT-Thresholdscan-Comparison-Latched-Noise}.

\begin{figure}[!tbp]\centering
\includegraphics[width=\textwidth]{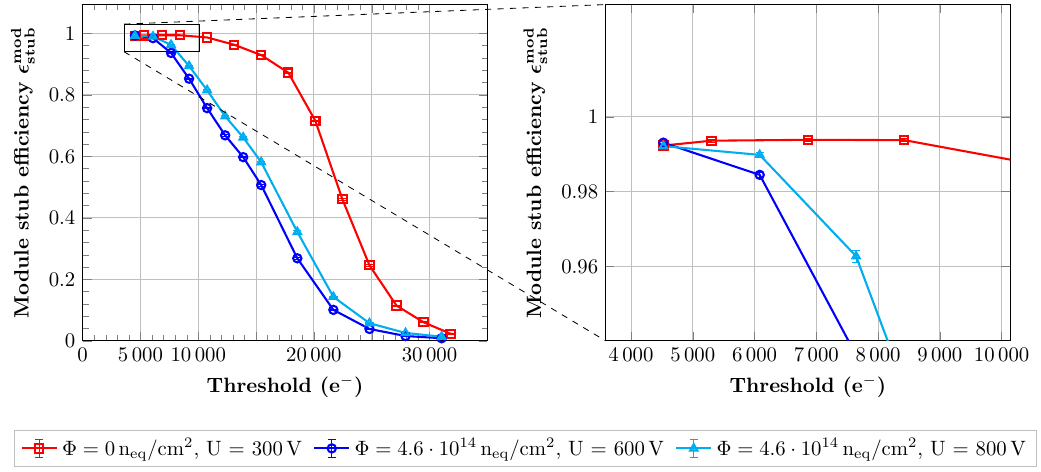}
\caption{Comparison of the module stub efficiency as a function of the threshold for module~5 (unirradiated, red markers) and module~4 (irradiated sensors, blue and cyan markers) at different sensor bias voltages.
The right plot shows a zoomed version of the left plot.
Statistical uncertainties are shown as error bars.
However, uncertainties are dominated by systematics, estimated in Section~\ref{sec:analysis}.}
\label{fig:KIT-StubEff-Thresholdscan-Comparison}
\end{figure}

Figure~\ref{fig:KIT-StubEff-Thresholdscan-Comparison} compares the module stub efficiency before and after irradiation.
As expected, the measured stub efficiency decreases with increasing threshold.
Both modules show a stub efficiency of about $\qty{99}{\percent}$ for low thresholds.
As in the case of the cluster efficiency, the stub efficiency is smaller after irradiation and starts decreasing at lower thresholds with respect to unirradiated sensors.
An increase in bias voltage from $\qty{600}{V}$ to $\qty{800}{V}$ yields a slightly higher stub efficiency at thresholds larger than $5000\,\textrm{e}^{-}$.

\begin{figure}[!tbp]\centering
    \includegraphics[width=\textwidth]{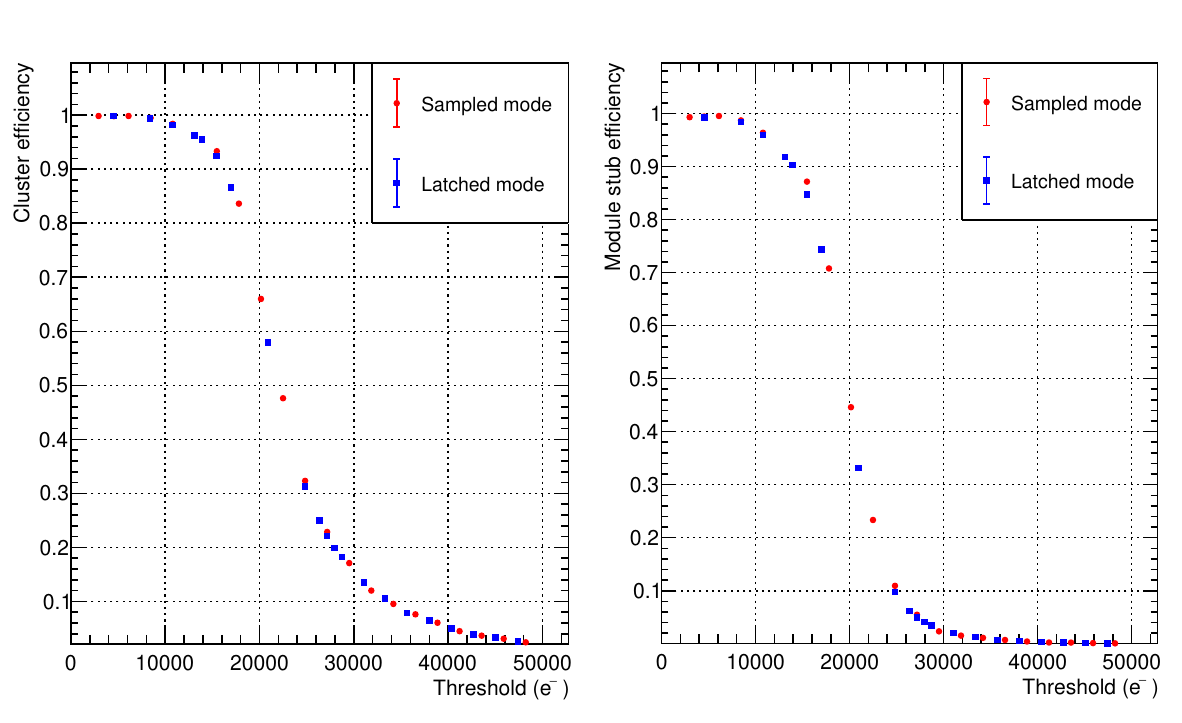}
    \caption[]{
    Cluster efficiencies of the bottom sensor layer (left) and module stub efficiencies (right) of the unirradiated module~1 in threshold scans for two hit detection modes, sampled mode and latched mode, at $\qty{300}{V}$ sensor bias.
    The measured efficiencies are shown as a function of the threshold in electrons.
    }
    \label{fig:thr-scan-bias-htd-comp}
\end{figure}

Figure~\ref{fig:thr-scan-bias-htd-comp} shows the comparison of threshold scans for the two tested hit detection modes, latched mode and sampled mode, for the unirradiated module~1 at a bias voltage of $\qty{300}{V}$.
The left side of Figure~\ref{fig:thr-scan-bias-htd-comp} shows the cluster efficiencies for the bottom sensor of the module.
The right side shows the module stub efficiencies.
As expected, no significant difference between the two hit detection modes is observed.

From the threshold scans, the signal amplitude distribution can be reconstructed by calculating the efficiency gradient as a function of the threshold.
The efficiency gradient $s_i$ can be calculated from the point-by-point gradient, using the threshold in electrons $\textrm{Thr}_i$
and the measured efficiency $\varepsilon_{\textrm{cluster},i}$ 
such that
\begin{align}
\label{eq:signal-spectrum}
s_{i}  &= - \frac{\varepsilon_{\textrm{cluster},i+1}-\varepsilon_{\textrm{cluster},i}}{\textrm{Thr}_{i+1}-\textrm{Thr}_{i}}~.
\end{align}

\begin{figure}[!tbp]\centering
    \includegraphics[width=0.75\textwidth]{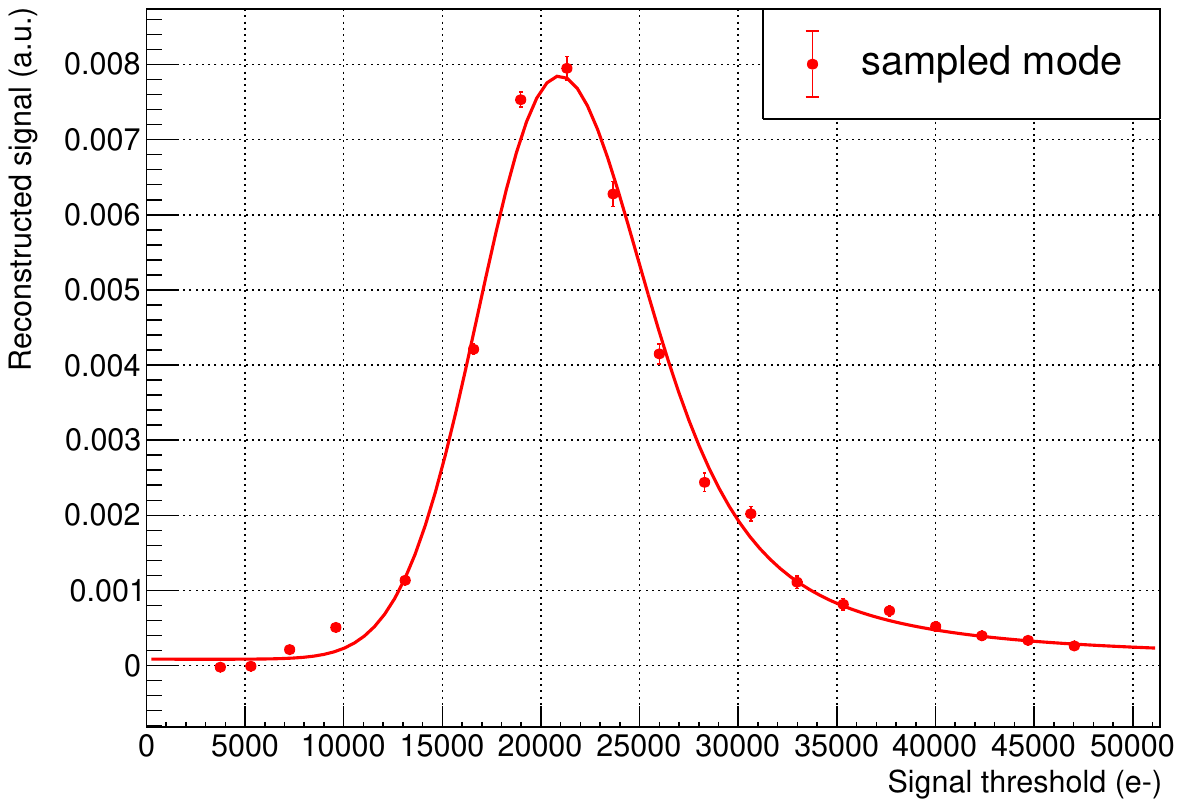}
    \caption[]{
    Reconstructed signal amplitude distribution from cluster efficiencies in threshold scans at $\qty{300}{V}$ sensor bias of the unirradiated module 1. 
    Data are shown as markers, while the fit is drawn as a solid line.}
    \label{fig:signal_reconstructed}
\end{figure}

The gradient as a function of the threshold is fitted with the convolution of a Landau function and a Gaussian function
\begin{equation}
    f(\textrm{Thr}_{i}, \mathbb{P}) = \left(\textrm{offset} + \textrm{scale} \times L(\textrm{Thr}_{i},{\textrm{MPV},\textrm{width}})\right) \circledast g(\textrm{Thr}_{i},\zeta)~,
    \label{eq:signal-fit-funct}
\end{equation}
where $\mathbb{P}$ is a parameter set consisting of an offset, a scale, the most probable value (MPV), a width, and a smearing parameter~$\zeta$.
The symbol~$L$ represents the Landau function and $g$ the Gaussian function.
This is an approximation to calculations and measurements shown in Ref.~\cite{Bak:1987cz}.
The model parameter~$\zeta$ is used because charged particles traversing thin silicon layers are not perfectly described by a Landau function alone.
Figure~\ref{fig:signal_reconstructed} shows the reconstructed signal for the threshold scan in sampled mode presented in Figure~\ref{fig:thr-scan-bias-htd-comp}.
The plot contains data acquired with the bottom sensor at $\qty{300}{V}$ with the corresponding Landau-Gaussian curve to guide the eye.
No further selection is applied on the hit positions and no particular cluster width is required.
Therefore, the signal amplitude distribution is expected to also contain the effects of charge sharing, which is expected to broaden the resulting distribution.
The distribution peaks at approximately $\num{21000}\,\textrm{e}^-$, comparable to previous results of $\num{21700}\,\textrm{e}^-$ (assuming a charge generation of $\qty{74.8}{e^{-}/\micro m}$ and a sensor thickness of $\qty{290}{\micro m}$)~\cite{Adam_2017}. 
No significant difference is observed between the data taken with the two different hit detection modes.
A comparison to module~4 with irradiated sensors was not possible as runs with this module did not cover a sufficiently large range in threshold values.

\subsection{Cluster Efficiency Across the Sensor Surface}

\begin{figure}[!tbp]\centering
    \includegraphics[width=\textwidth]{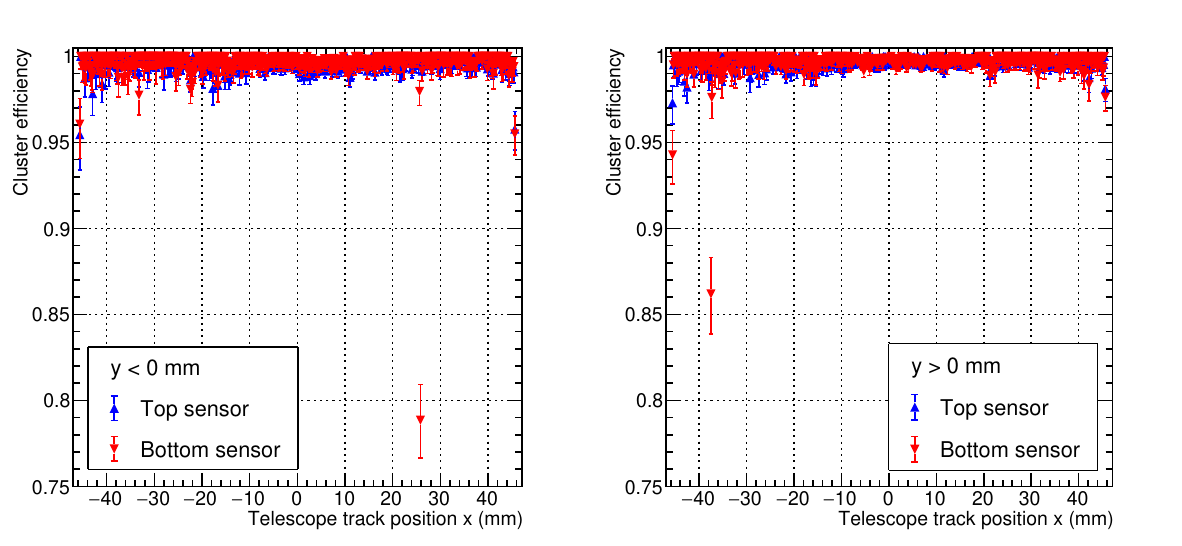} 
\caption{Cluster efficiency of module~1 as a function of the $x$-position for the top (blue) and bottom (red) sensor. The left plot shows the region of $y<\qty{0}{\milli\meter}$, the right plot the region of $y>\qty{0}{\milli\meter}$. The bin width in $x$ is $\qty{180}{\micro\meter}$ (twice the strip pitch). The given uncertainties are statistical only. The measurements are performed with a threshold of approximately $\num{6000}\,\textrm{e}^{-}$.}
\label{fig:hiteffx}
\end{figure}

The cluster efficiency along the track position in $x$ is shown in Figure~\ref{fig:hiteffx} for module~1.
The mean cluster efficiency is $\qty{99.75}{\percent}$ over the entire tested area on both sensors, including two bins with lower efficiency that correspond to two channels on the bottom sensor which are likely disconnected from the readout electronics, and the bins at the edges of the sensor.
The fact that the cluster efficiency is still above $\qty{75}{\percent}$ for the presumed disconnected strips is related to histogram binning effects, the cluster efficiency distance criteria and track pointing uncertainties.

\begin{figure}[!tbp]\centering
    \includegraphics[width=0.5\textwidth]{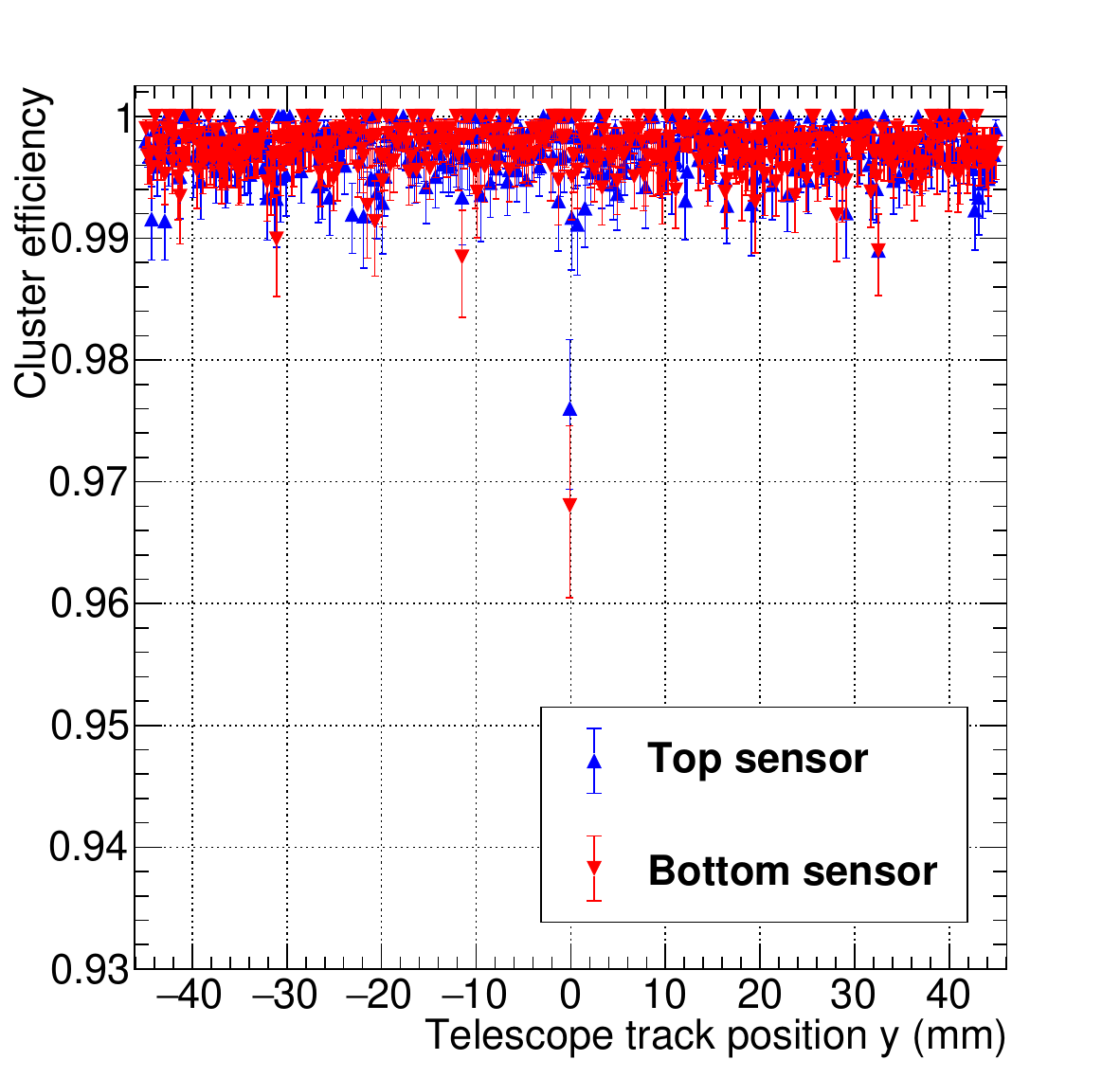}
	\caption{Cluster efficiency of module~1 as a function of the $y$-position for the top (blue) and bottom (red) sensor.
 The bin width in $y$ is $\qty{200}{\micro\meter}$.
 The shown uncertainties are statistical only. The measurements are done with a hit detection threshold of approximately $\num{6000}\,\textrm{e}^-$.}
\label{fig:hiteffy}
\end{figure}

\begin{figure}[!tbp]\centering
    \includegraphics[width=0.6\textwidth]{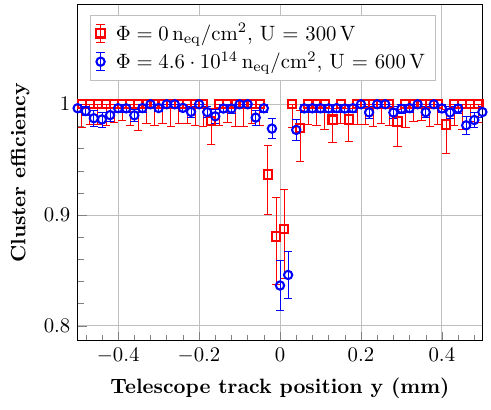}
	\caption{Comparison of the cluster efficiency as a function of the $y$-position in the center of the sensor for module~1 (unirradiated, red markers) and module~4 (irradiated, blue markers).
    The threshold is set to approximately $\num{6000}\,\textrm{e}^-$ for both data sets.
    Only data from the bottom sensors are shown.
    Data from the top sensor show similar results. 
    The binning in the $y$-position is $\qty{20}{\micro\meter}$.
    }
    \label{fig:hiteffy_irrad}
\end{figure}

The cluster efficiency as a function of the $y$-position is shown in Figure~\ref{fig:hiteffy} for module~1.
In the center region of the module ($y \approx \qty{0}{mm}$), where the two rows of strips meet, the efficiency is decreased by approximately $\qty{2}{\percent}$ over the bin region of $\qty{200}{\micron}$.
To compare this effect before and after irradiation, Figure~\ref{fig:hiteffy_irrad} shows the cluster efficiency in the area $-\qty{0.5}{mm} \leq y \leq +\qty{0.5}{mm}$ for modules~1 and 4 with a bin width of $\qty{20}{\micro\meter}$.
Both sets of data have been taken at a threshold of approximately $\num{6000}\,\textrm{e}^-$.
For this threshold setting, no significant differences in the cluster efficiency in the sensor middle can be observed before and after irradiation at the respective nominal bias voltage settings of \qty{300}{V} and \qty{600}{V}, respectively.

\begin{figure}[!tbp]\centering
\includegraphics[width=0.67\textwidth]{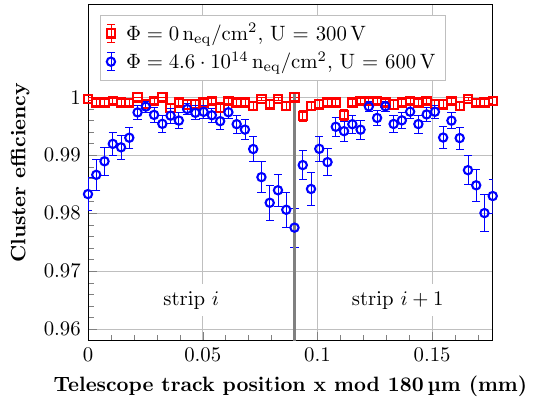}
\caption{Measured cluster efficiency within two neighboring strips before irradiation (module~5, red data points) and after irradiation (module~4, blue data points).
Only data from the bottom sensors are shown.
Data from the top sensors are identical.
The threshold is set to approximately $\num{6000}\,\textrm{e}^-$ for both data sets.
To reduce statistical uncertainties, information on all illuminated strips is folded into the $x$ coordinates of two neighboring strips.}
\label{fig:KIT-Inpixelefficiency}
\end{figure}

Figure~\ref{fig:KIT-Inpixelefficiency} shows the cluster efficiency distribution of the irradiated module~4 and the unirradiated module~5 within two neighboring strips. 
The information of all illuminated strips are folded into the plot by performing the modulo operation of the reconstructed $x$ coordinate of the track on the DUT with twice the strip pitch. 
While the unirradiated sensor shows a constant cluster efficiency over the entire strip width, the irradiated sensor shows a maximum cluster efficiency directly below the strip implant.
Between the strips, the cluster efficiency decreases by $\qty{2}{\percent}$. 
The combination of smaller signals after irradiation and the increased probability for charge sharing in the region between strips leads to this efficiency decrease. 

\subsection{Performance of Transverse Momentum Discrimination}\label{subsec:stub_efficiency_transverse_momentum}
\begin{figure}[!tbp]
\centering
\includegraphics[width=0.9\textwidth]{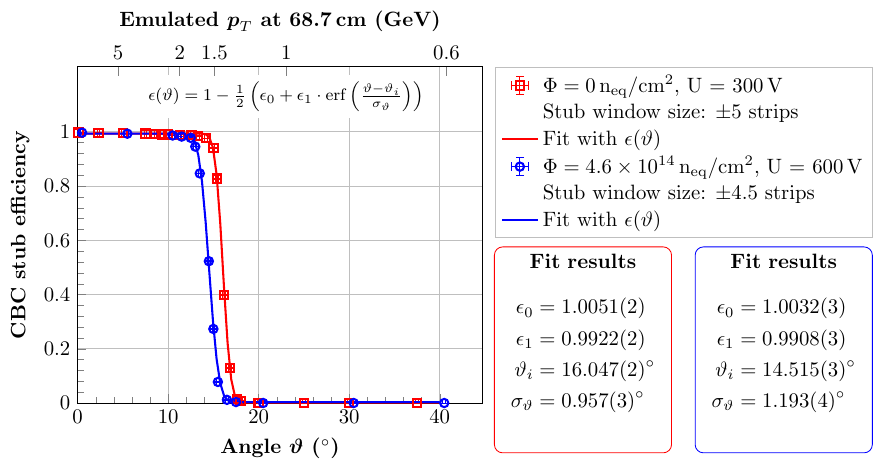}
    \caption{CBC stub efficiency as a function of the rotation angle $\vartheta$. Data from the unirradiated module (module~1) are shown in red and data from the irradiated module (module~4) are shown in blue. Two different stub window sizes have been used for the measurements.
    The top $x$-axis indicates the emulated transverse momentum for a 2S module in the CMS tracker at a radial position of \qty{68.7}{cm}. 
    Statistical uncertainties are shown as error bars.
    }
    \label{fig:CBCstubefficiency_angle}
\end{figure}

By turning the module around the axis parallel to the strip orientation, it is possible to emulate different \pt~values to investigate the performance of the stub-finding logic.
Figure~\ref{fig:CBCstubefficiency_angle} shows the CBC stub efficiencies as a function of the module turning angle~$\vartheta$ for modules with unirradiated and irradiated sensors at a threshold of approximately $\num{6000}\,\textrm{e}^-$.
Two different stub window sizes were used during two different beam tests. 
The stub window size was set to $\pm 5$~strips for measurements with module~1 and to $\pm 4.5$~strips for module~4 with irradiated sensors.
For small rotation angles, the particle tracks generate clusters in the correlation layer within the stub window.
Thus, stubs are produced and the CBC stub efficiency is constant at $\qty{100}{\percent}$.
By increasing the rotation angle, the cluster center in the correlation layer moves toward the edge of the stub window.
This results in a drop of the stub efficiency to zero.
A larger stub window size leads to a drop of the stub efficiency at higher rotation angles.

To quantify the measured stub efficiency distribution, an error function of the form
\begin{equation}
\epsilon_{\textrm{stub}}^{\textrm{CBC}}(\vartheta) = 1 - \frac{1}{2} \left(\epsilon_0 + \epsilon_1 \cdot \textrm{erf}\left(\frac{\vartheta - \vartheta_i}{\sigma_\vartheta}\right)\right)
\label{eq:stubeff_turnon}
\end{equation}
is used.
The parameters $\epsilon_0$ and $\epsilon_1$ describe the vertical position and the scaling of the distribution and are expected to be compatible with~$\num{1}$.
The parameter $\vartheta_i$ indicates the angular position of the inflection point and $\sigma_\vartheta$ quantifies the width of the distribution.
The fit values are summarized in Figure~\ref{fig:CBCstubefficiency_angle}.
For the analysis, statistical and systematic uncertainties are included.
The stub efficiency distribution is sensitive to misalignments of the DUT rotation stage around the $y$ axis during the beam test and the relative sensor alignment in the {2S~modules}.
Following the studies presented in Ref.~\cite{PhdRolandKoppenhoefer}, a combined systematic uncertainty of $0.15^\circ$ is used for the analysis.

In the case of module~1 measured with a stub window size of $\pm 5$~strips the drop is observed at an angle of around $16^\circ$ while for module~4 with a stub window size of $\pm 4.5$~strips the drop occurs at around $14.5^\circ$.
As discussed in Ref.~\cite{PhdRolandKoppenhoefer}, the distance between top and bottom sensor in the investigated 2S module prototypes is not constant across each module but reduces towards the modules' center. 
Considering this effect and the different chosen stub window sizes for the measurements, the extracted positions of the drop in the stub efficiency distributions are in agreement with the geometrical expectation. 

Within a magnetic field charged particles with transverse momentum \pt{} are forced onto a circular trajectory in the plane perpendicular to the beam axis.
The radius of the trajectory is given by
\begin{equation}
\label{eq:rTpTB}
r_{\text{T}} [\text{m}] = \frac{p_{\textrm{T}} [\text{GeV}]}{0.3 \cdot B[\text{T}]}~.
\end{equation}
The magnetic field strength present in the CMS tracker is $B = \qty{3.8}{T}$. 
Together with Equation~\ref{eq:rTpTB} the track incidence angle $\vartheta$ for {2S~modules} at a radial position $R$ can be converted into an emulated transverse momentum using
\begin{equation}
\label{eq:anglePt}
p_{\textrm{T}} [\text{GeV}] = \frac{0.57 \cdot R[\text{m}]}{\sin{\vartheta}}~.
\end{equation}
Equation~\ref{eq:anglePt} is evaluated in this paper for \mbox{$R = \qty{68.7}{cm}$}, corresponding to the smallest distance of {2S~modules} to the interaction point.
The top $x$-axis in Figure~\ref{fig:CBCstubefficiency_angle} shows the corresponding \pt~values. 
By choosing a smaller stub window size, a tighter transverse momentum cut, i.e. at a higher \pt~value, can be achieved.
The unirradiated and irradiated sensors show a sharp turn-on of the stub efficiency at a \pt~threshold of $\qty{1.4}{GeV}$ and $\qty{1.6}{GeV}$, respectively.

For transverse momenta, the stub efficiency curve can be described with an error function of the form
\begin{equation}
  \label{eq:error_pt}
  \epsilon_{\textrm{stub}}^{\textrm{CBC}}(p_{\text{T}})  = \frac{1}{2} \left( \Tilde{\epsilon_0} + \Tilde{\epsilon_1} \cdot \textrm{erf}\left(\frac{p_{\text{T}} - p_i}{\sigma_{p_{\text{T}}}}\right)\right)~.
\end{equation}
The relative transverse momentum resolution can then be defined as $\frac{\sigma_{p_{\text{T}}}}{p_i}$.
Resolutions of $\qty{5.8}{\percent}$ and $\qty{8.1}{\percent}$ are resulting for the unirradiated module~1 and module~4 with irradiated sensors, respectively.
However, a conclusive comparison of these values is not possible based on the available data sets from the beam test campaigns investigated.
The value of $\sigma_{p_{\text{T}}}$ depends on several factors such as e.g.\ the correlation window size, the incidence angle distribution of tracks onto the DUT at the beam line, or the observed non-constant spacing of the two sensors in each module~\cite{PhdRolandKoppenhoefer, PhdTimZiemons}.
As mentioned above, the presented data sets have been gathered with two different modules at different window size settings and in two different beam lines.
A direct comparison of the transverse momentum discrimination performance is therefore only possible with future data.

\begin{figure}[!tbp]\centering
    \includegraphics[width=0.65\textwidth]{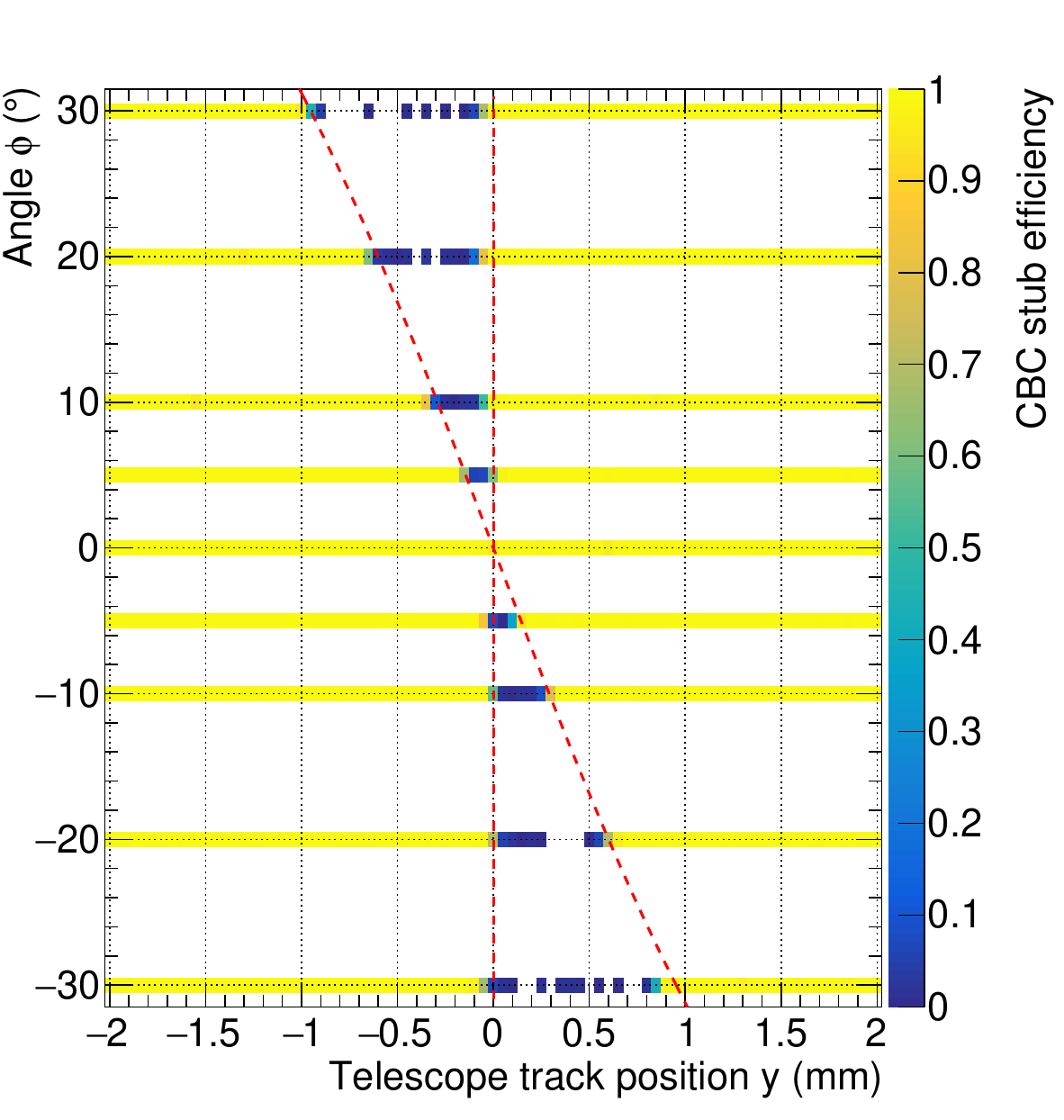}
    \caption{{CBC} stub efficiency as a function of the incident angle $\phi$ and the track $y$-position shown for the unirradiated module~1. The module has been rotated around the mid-sensor axis ($x$-axis). The red dashed lines show the expected borders of the stub-insensitive region assuming a distance between the top and the bottom sensor of $\qty{1.65}{mm}$.}
\label{fig:InclinationScan2DStubEfficiency}
\end{figure}

The stub information is produced by the CBCs on the FEHs that read out only one of the two sensor halves.
Stubs are formed from hit information on both the top and the bottom sensors.
Therefore, the {2S~module} is expected to have an inefficiency in stub detection at the center of the sensor ($y \approx \qty{0}{\milli\meter}$) if the particle is detected by different FEHs on the different sensor halves.
To confirm this, the unirradiated module~1 is rotated along the mid-sensor axis (along $x$ at $y=\qty{0}{\milli\meter}$) perpendicular to the strips.
When the angle of incidence of the particles is increased, the width of the inefficient area is expected to increase.
Assuming a simple geometric approach, the width of the inefficient area, $W$, can be calculated by
$W = d \cdot \left\lvert \tan \left( \phi \right) \right\rvert$,
where $d$ is the distance between the top and bottom sensors and $\phi$ is the angle of incidence.
Figure~\ref{fig:InclinationScan2DStubEfficiency} shows the {CBC} stub efficiency as a function of the incidence angle and the track $y$-coordinate of the unirradiated module~1.
As expected, the stub efficiency drops in the central region of the DUT and
the inefficient region becomes larger as the absolute angle of incidence is increased. 
The inefficient area is not perfectly located in the center at $y=\qty{0}{mm}$ because of uncertainties in the alignment of the $y$-coordinate within the telescope and because the axis of rotation is not exactly in the middle between the two sensors.
The width of the inefficient area is in good agreement with the simple geometric approach assuming a sensor distance of approximately $\qty{1.65}{mm}$.
The discussed effect is relevant as {2S~modules} will be installed in the OT barrel at about $r=\qty{68.7}{cm}$ and up to $z=\qty{117.6}{cm}$ leading to incidence angles of up to $\qty{60}{\degree}$. 
With a sensor length of approximately $\qty{10}{cm}$ up to 
$({\qty{1.8}{mm}}\cdot \tan \qty{60}{\degree})/\qty{100}{mm} \approx \qty{3}{\percent}$
of the module surface will be inefficient to stubs, depending on the module position in the tracker.

\begin{figure}[!tbp]
    \centering
      \includegraphics[width=0.8\textwidth]{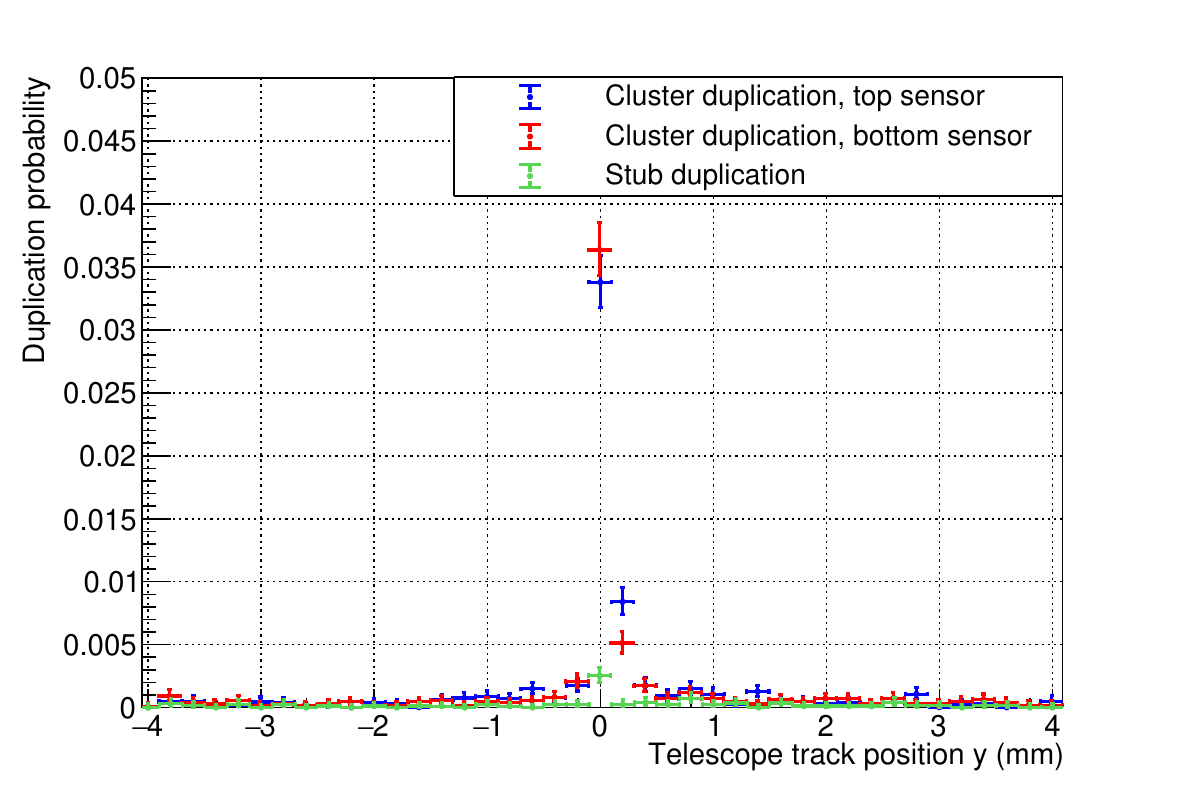}
      \caption{Cluster duplication probability (bottom sensor in red, top sensor in blue) and stub duplication probability (in green) of module~5 at different positions of $y$.
      The measurement has been taken at a bias voltage of $\qty{600}{\volt}$.
      The coordinate $y=\qty{0}{\milli\meter}$ represents the boundary between the two sensor halves.
      The binning is chosen to be $\qty{200}{\micro\meter}$ in $y$.}
    \label{fig:DuplicationProbability}
\end{figure}

Hits in the center region of the sensors could be detected on both sensor sides by charge sharing along the strip direction.
Figure~\ref{fig:DuplicationProbability} presents the cluster and stub duplication probability along the strip axis ($y$-axis) for module~5 at $\qty{600}{V}$ bias voltage and a threshold of approximately $\num{5000}\,\textrm{e}^-$ at perpendicular particle incidence.
The cluster duplication probability is determined by correlating the number of tracks linked to clusters with a size of two strips in the $y$ direction (i.e.\ hits in both strips on either side of the module center) to the total number of tracks linked to any cluster.
Only tracks fulfilling the cluster efficiency criterion are considered.
The number of duplicated stub candidates is the number with tracks that lead to clusters in both sensor halves, and that are linked to one stub on each FEH of $\Delta x_\textrm{stubs} < \qty{300}{\micro\meter}$. 
By comparing this number to the number of all tracks linked to a cluster on both sensor layers, the stub duplication probability is determined.
At the center of the module ($-\qty{200}{\micro\meter} < y < +\qty{200}{\micro\meter}$) the cluster duplication probability is measured to be approximately $\qty{3.5}{\percent}$ and the stub duplication probability is found to be about $\qty{0.3}{\percent}$.
Given the small area in which duplication takes place, the impact on readout bandwidth and overall tracking performance in CMS is expected to be negligible. 

\subsection{Charge Sharing}
Charge-sharing effects are investigated by measuring the cluster size under varying incidence angles.
The mean cluster size as a function of the incidence angle can be modeled by the geometric relation \begin{equation}
\label{eq:clusterSize}
s(\vartheta) = s_0 + \frac{t}{p} \cdot \lvert \tan{ (\vartheta - \vartheta_0) \rvert } ~,
\end{equation}
where $s$ is the mean cluster size, $s_0$ the mean cluster size at zero incident angle, $t$ the active thickness of the sensor, $p$ the pitch of the strips, $\vartheta$ the incident angle given as the rotation stage setting and $\vartheta_0$ the difference between the setting of the rotation stage and the real incident angle.
Dependencies on the threshold and inhomogeneities of the electric field inside the sensors are included by replacing the fraction of fixed sensor geometry parameters $\frac{t}{p}$ by a free parameter $\kappa$.
Additionally, Equation~\ref{eq:clusterSize} is convolved with a Gaussian function with width $\sigma_\textrm{u}$ to account for diffusion effects and angular uncertainties~\cite{Haranko:2020vlj}, 
resulting in
\begin{equation}
\label{eq:clusterSizeParametrization}
s \left( \vartheta \right) = \frac{1}{\sqrt{2 \pi} \sigma_\textrm{u}} \cdot \int_{-\infty}^{\infty} \left[ s_0 + \kappa \lvert \tan{ \left( \vartheta'- \vartheta_0 \right) } \rvert \right] \exp{ \left(-\frac{ \left( \vartheta'-\vartheta \right)^2}{2 \sigma_\textrm{u}^2} \right) } \,d\vartheta' ~.
\end{equation}

\begin{figure}[!tbp]\centering
    \includegraphics[width=0.6\textwidth]{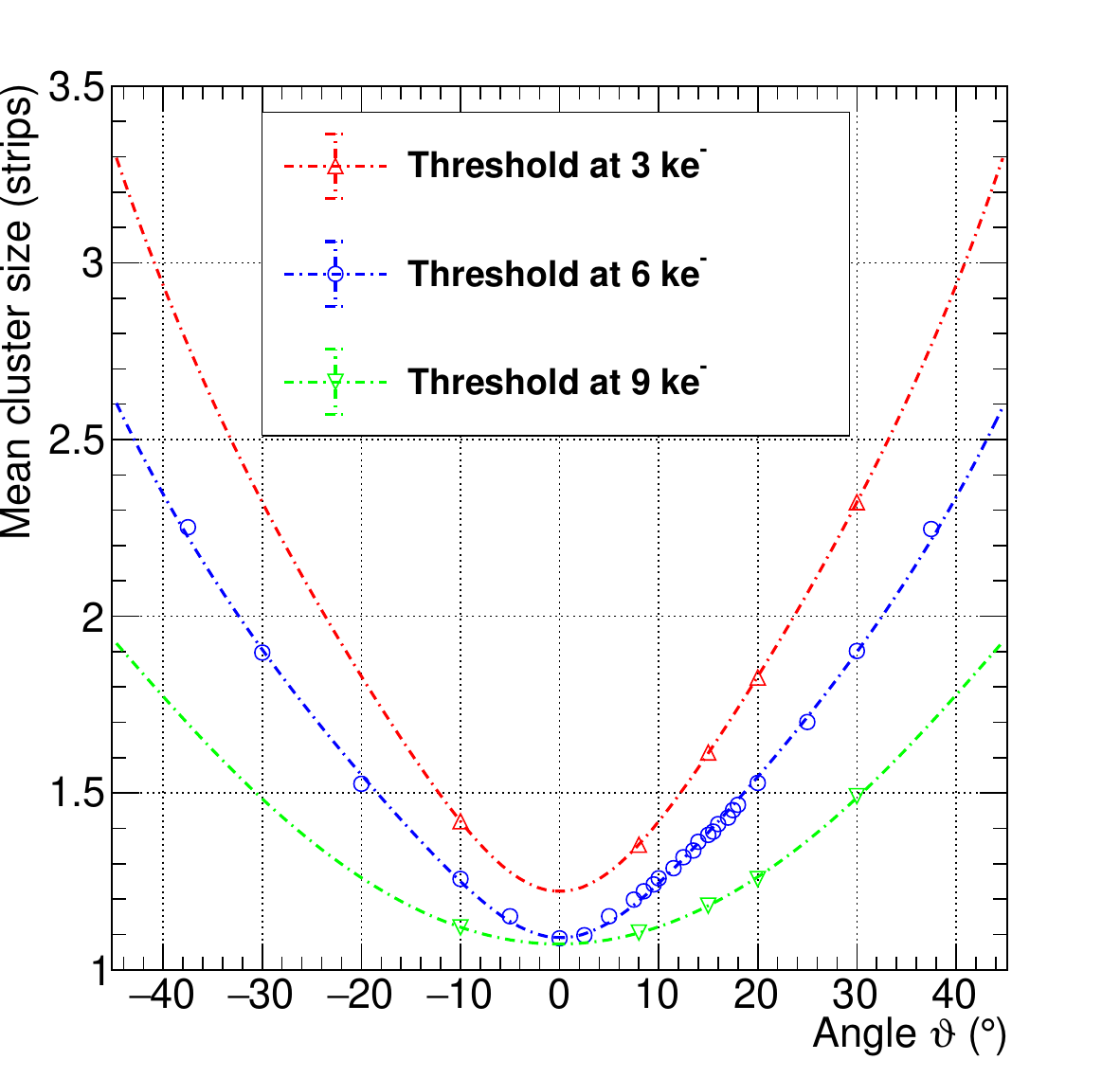}
    \caption{Arithmetic mean of the cluster size in the top and bottom sensor as a function of the incidence angle on the unirradiated module~1 for different threshold settings.}
    \label{fig:cluswidthvsthreshold}
\end{figure}

Figure~\ref{fig:cluswidthvsthreshold} shows the mean cluster size of the top and bottom sensor as a function of the incidence angle for different threshold values for the unirradiated module~1.
The measurements are well represented by the model described above and the fit parameters are summarized in Table~\ref{tab:fit-param-cluster-width}.
As expected, the fit parameter $\kappa$ increases almost linearly with decreasing threshold.
The fit results can be used to tune the \GEANTfour-based full detector simulation of the CMS detector.

\begin{table}[!tb]\centering
    \caption[]{Summary of the fit parameters of the mean cluster size as a function of the incidence angle.}
    \label{tab:fit-param-cluster-width}
    \vspace{0.2cm }
    \begin{tabular}{lcccc}
    \hline
      Threshold & $\vartheta_{0} (^\circ)$ & $s_{0}$ & $\sigma_\textrm{u} (^\circ)$ & $\kappa$            \\
      \hline
      $\num{3000}~\textrm{e}^-$        & $0.01 \pm 0.10$     & $1.00$  & $7.1  \pm 0.7$   & $2.25 \pm 0.03$ \\
      $\num{6000}~\textrm{e}^-$              & $0.10 \pm 0.04$     & $0.95$  & $6.2  \pm 0.1$   & $1.62 \pm 0.01$  \\
      $\num{9000}~\textrm{e}^-$           & $-0.05 \pm 0.12$     & $0.77$  & $19.4 \pm 1.9$   & $1.01 \pm 0.03$  \\
      \hline
      \end{tabular}
\end{table}

\begin{figure}[!tbp]\centering
    \includegraphics[width=.95\textwidth]{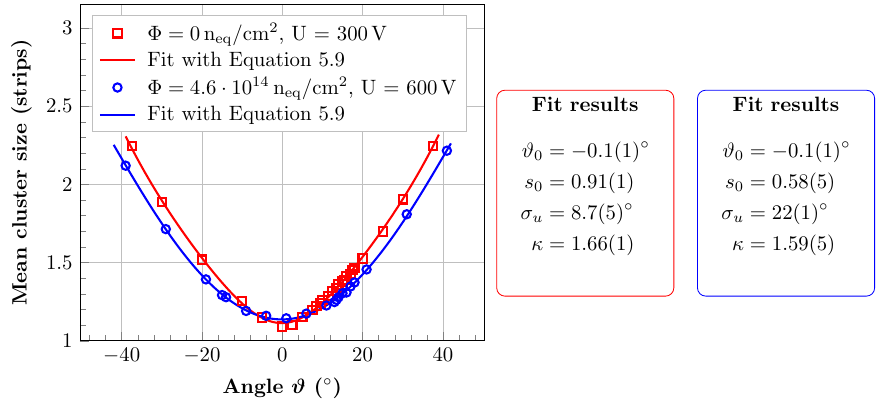}
    \caption{Comparison of the mean cluster size as a function of the incidence angle for the unirradiated module~1 (red data points) and for the irradiated module~4 (blue data points). The threshold is set to $\num{6000}\,\textrm{e}^-$. The red and blue lines indicate the corresponding fit results using Equation~\ref{eq:clusterSizeParametrization}. }
    \label{fig:cluswidthvsthreshold-irrad}
\end{figure}

Figure~\ref{fig:cluswidthvsthreshold-irrad} compares the measured mean cluster sizes before and after irradiation (module~1 and 4, respectively) at a similar threshold.
For zero incidence angle the modules show comparable mean cluster sizes.
For increasing angles, the mean cluster sizes after irradiation are smaller than before irradiation.
This is consistent with the expectation that the cluster charge of the sensors will decrease after irradiation due to the induced radiation damage.
At high incidence angles, the sensor signal is shared between several strips and the probability that the signal drops below the threshold increases with irradiation.

\begin{figure}[!tbp]\centering
      \includegraphics[width=.7\textwidth]{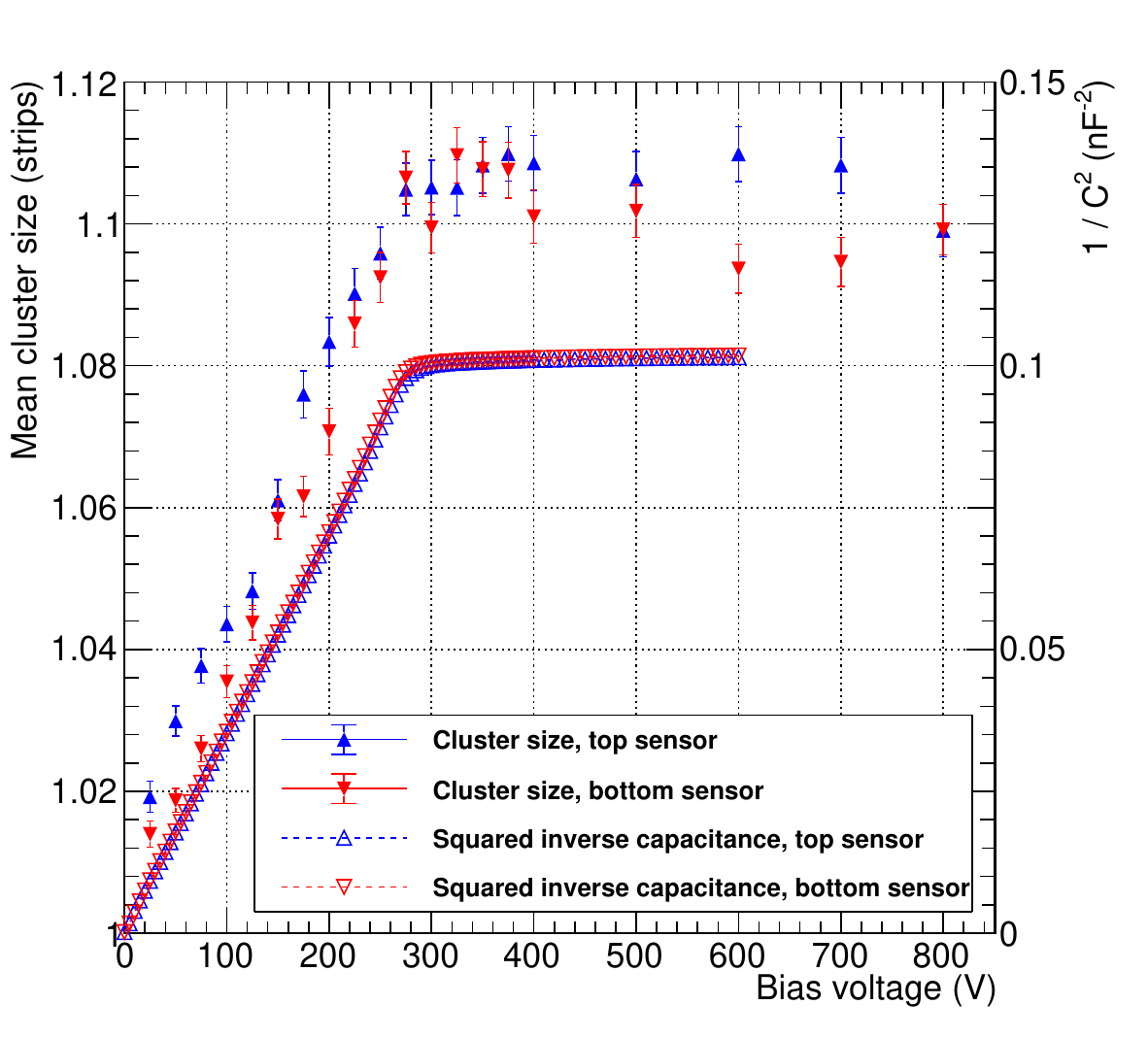}
      \caption{Mean cluster size in the top (blue markers) and bottom sensor (red markers) of module~5 as a function of the bias voltage. The threshold is set to about $\num{5000}\,\textrm{e}^-$. The shown error bars present the statistical uncertainty. The capacitance measurement for both sensors is shown in the same plot with the scale on the right.}
      \label{fig:biasscanclustersize}
\end{figure}

Figure~\ref{fig:biasscanclustersize} shows the mean cluster size as a function of the bias voltage measured using module~5.
The mean cluster size of both sensors increases with bias voltage up to a value of approximately $\qty{275}{V}$.
Full depletion is reached at this bias voltage.
The capacitance measurement taken in the laboratory before the beam test is shown in the same figure as the squared inverse and its plateau is reached at the same bias voltage, confirming the extracted full depletion voltage of the sensors.

\subsection{Readout Stability and Correlations between Modules}
\begin{figure}[!tbp]
    \centering
    \includegraphics[width=0.6\textwidth]{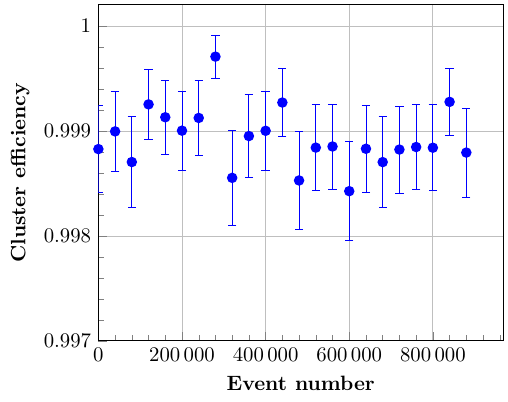} 
    \caption[]{Averaged cluster efficiency as a function of the event number. Data acquisition took about $20$~minutes.
    Error bars indicate statistical uncertainties. }
    \label{fig:simuRO-event-effi}
\end{figure}

Other important aspects of the beam test campaigns were the studies of synchronous readout of two modules using a single FC7 board and of the readout stability.
To assess the stability of the optical readout, Figure~\ref{fig:simuRO-event-effi} shows the cluster efficiency for the unirradiated module~5 over the course of the event readout, indicated by the increasing event number. 
No discernible time dependence is observed over $20$~minutes in this example.

\begin{figure}[!tbp]
    \centering
    \includegraphics[width=0.8\textwidth]{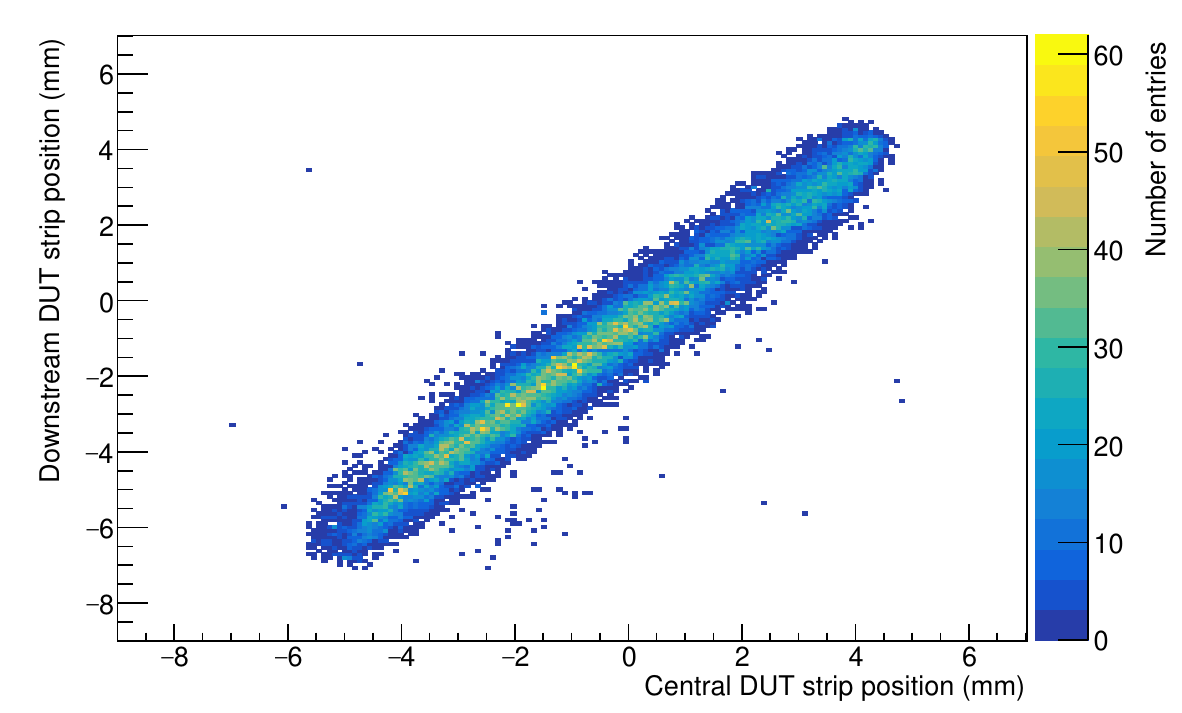}
    \caption[Stub-to-stub correlation]{
    Stub position correlation of the central module~1 and the downstream module~3 for selected events. The number of measured stubs is coded in the color scale.
    }
    \label{fig:simuRO-hitCorrel-stub-2}
\end{figure}

A clear stub correlation between two modules is shown in Figure~\ref{fig:simuRO-hitCorrel-stub-2}, showing the stub data of the four FEHs of modules~1 and 3.
As stubs and hits are evaluated separately in the software, the tracks shown here are linked to the selected stubs and also to pairs of efficient clusters as defined in Section~\ref{sec:analysis}.
No requirement on the track straightness is made.
Stubs measured on module~1 are correlated with stubs on module~3 using the modules' local coordinate frames.
Very few additional correlations are visible outside the beam spot region, possibly due to larger scattering angles of the electron beam in the telescope.
The correlation band has a width of approximately $\qty{550}{\micron}$ in the stub-to-stub position residual.
The distance from module~1 to module~3 is approximately $\qty{530}{\milli\meter}$, the resulting mean angle of deflection is $\qty{0.06}{\degree}$.
A result of the same order of magnitude is obtained when estimating the multiple scattering angle due to all material present between module~1 and module~3.

%% file: DN-20-014_conclusion.tex
\section{Conclusions} \label{sec:conclusion}
The CMS tracker group performed beam test campaigns with electron beams at the {DESY-II} Test Beam Facility in Hamburg on multiple full-size 2S module prototypes developed for the Outer Tracker of CMS at the HL-LHC.

One of the modules has been built using sensors irradiated with protons to a fluence of $\num{4.6E14}\,\textrm{n}_{\textrm{eq}}\textrm{cm}^{-2}$, corresponding to $\qty{91}{\percent}$ of the maximum expected fluence for 2S modules after $\qty{4000}{fb}^{-1}$.
The performance of the modules has been evaluated under various operating conditions and several aspects have been studied in detail.
There is no significant deviation from the results of previous test beam studies and expectations from the module geometry.

Modules with unirradiated sensors can be operated 
over a large threshold range and reach a cluster efficiency larger than $\qty{99.5}{\percent}$.
For the irradiated sensors 
at the nominal operating voltage of $\qty{600}{V}$, 
the maximum hit efficiency is slightly lower than in the unirradiated case, but still well above $\qty{99}{\percent}$.
By increasing the sensor bias voltage to $\qty{800}{V}$ or by reducing the threshold the loss can be recovered partially, an option under consideration for the modules with the highest expected fluences in the tracker.
No dependency of the efficiency for different positions on the module surface other than the central region has been found.
This shows that the sensors of the 2S modules can be operated with high efficiency, low noise occupancy and with the expected spatial resolution throughout the lifetime of the tracker at the HL-LHC.

The functionality of the \pt{} discrimination logic of the CBC chip was studied in detail.
For both the unirradiated modules and the modules with irradiated sensors, a module stub efficiency of about $\qty{99}{\percent}$ at thresholds below $\num{6000}\,\textrm{e}^-$ was achieved for particles traversing the module surface perpendicularly.
The suppression of low-\pt{} particles was tested by rotating the module along the axis parallel to the strips.
The performance of the \pt{} discrimination meets the geometric expectations, demonstrating the correct functioning of the stub finding logic.
No significant degradation of the stub performance has been found for the module with irradiated sensors. 

The performance of the prototype 2S modules obtained in these beam test campaigns at the DESY electron beam, also with irradiated sensors, meets the requirements for operation in the CMS Outer Tracker at the HL-LHC.

%% file: TrackerAuthorList_2024_OT2STestbeam.tex
\setlength{\parindent}{0pt}

\newcommand{\cmsAuthorMark}[1]
{\hbox{\textsuperscript{\normalfont#1}}}

{\Large \textbf{The Tracker Group of the CMS Collaboration}\par}

\vspace{1cm}

\textcolor{black}{\textbf{Institut~f\"{u}r~Hochenergiephysik, Wien, Austria}\\*[0pt]
W.~Adam, T.~Bergauer, K.~Damanakis, M.~Dragicevic, R.~Fr\"{u}hwirth\cmsAuthorMark{1}, H.~Steininger}
\\

\textcolor{black}{\textbf{Universiteit~Antwerpen, Antwerpen, Belgium}\\*[0pt]
W.~Beaumont, M.R.~Darwish\cmsAuthorMark{2,3}, T.~Janssen, P.~Van~Mechelen
% ...
}
\\

\textcolor{black}{\textbf{Texas~A\&M~University, College~Station, USA}\\*[0pt]
R.~Eusebi}
\\

\textcolor{black}{\textbf{Vanderbilt~University, Nashville, USA}\\*[0pt]
P.~D'Angelo, W.~Johns}
\\

% ...
1: Also at Vienna University of Technology, Vienna, Austria \\
2: Also at Institute of Basic and Applied Sciences, Faculty of Engineering, Arab Academy for Science, Technology and Maritime Transport, Alexandria, Egypt \\
3: Now at Baylor University, Waco, USA\\
% ...
